\documentclass[aps,11pt,prd,showpacs]{revtex4-1}
\usepackage[utf8x]{inputenc}
\usepackage{amsmath} % Enhanced mathematics
\usepackage{amssymb} % Enhanced mathematics
\usepackage{graphics}
\usepackage{epsfig}
\usepackage{slashed}

\newcommand\ba{\begin{eqnarray}}
\newcommand\ea{\end{eqnarray}}
\newcommand\alb{\begin{align}}
\newcommand\ale{\end{align}}
\newcommand\be{\begin{equation}}
\newcommand\ee{\end{equation}}

\usepackage{longtable}
\usepackage{latexsym}

\usepackage{amssymb}
\usepackage{amsfonts} % for \mathfrak
\usepackage{amsmath}

\begin{document}

\title{ABOUT  THE POLARIZED $\tau$ LEPTON RADIATIVE DECAY.}

\vspace{4mm}

\author{G.I.~Gakh}

\affiliation{\it NSC ''Kharkov Institute of Physics and Technology'',
Akademicheskaya, 1, 61108 Kharkov,  and \\
V.N.~Karazin Kharkiv National University, 61022 Kharkov, Ukraine}

\author{M.I.~Konchatnij}
\affiliation{\it NSC ''Kharkov Institute of Physics and Technology'',
Akademicheskaya, 1, 61108 Kharkov, and \\
V.N.~Karazin Kharkiv National University, 61022 Kharkov, Ukraine}

\author{N.P.~ Merenkov}
\affiliation{\it NSC ''Kharkov Institute of Physics and Technology'',
Akademicheskaya, 1, 61108 Kharkov,  and \\
 V.N.~Karazin Kharkiv National University, 61022 Kharkov, Ukraine}

\date{\today}

\begin{abstract}
 The polarization effects in the one-meson
radiative decay of the polarized $\tau$ lepton,
$\tau^-\to\pi^-\gamma\nu_{\tau},$
%and $\tau^-\to K^-\gamma\nu_{\tau}$,
have been investigated.
We present the analytical results for the $t-$distribution of the partial differential widths,
which responsible for different polarization phenomena, in the case of the photon energy cut: $\omega>X$.
These quantities depend on the invariant mass squared $t$ of the pseudoscalar meson and photon and the photon energy cut $X.$
Our analytical formulae, in terms of the weak vector and axial-vector form factors, describing the structure-dependent part
of the decay amplitude, are valid also for the decay $\tau^{\pm}\to K^{\pm}\,\nu_\tau\,\gamma$
after trivial substitutions of the corresponding constants.
We demonstrate the essential decrease of the inner bremsstrahlung contribution in comparison with the structural one
with increase of the photon cut energy. In numerical calculations the vector and
axial-vector form factors are determined using the chiral effective
theory with resonances (R$\chi$T).

%The inner
%bremsstrahlung and structural amplitudes were taken into account.
%The asymmetry of the differential decay width caused by the $\tau$
%lepton polarization and the Stokes parameters of the emitted
%photon itself are calculated depending on the polarization of the decaying
%$\tau$ lepton. These
%physical quantities are estimated numerically for an arbitrary direction of the $\tau$
%lepton polarization 3-vector in the rest frame.
\end{abstract}

\maketitle

\section{Introduction}

\hspace{0.7cm}

The $\tau-$lepton physics attracts a large attention of both, theoreticians and experimentalists, from the day of its discovery up to nowadays.
The review of the different aspects of $\tau$ physics can be found in Refs. ~\cite{P13, N14}~. In the last years, the interest to different $\tau$ decays is stimulated by the plans to construct and run Super KEKB (Japan), Super $c-\tau$ (Russia) and Super $c-\tau$ (HIEPA, China)  facilities \cite{Z11,Lev08,O09,hiepa}~, where it will be accumulated more than 10$^{10}$ events with
polarized $\tau,$ and the modern status of these facilities is discussed in Refs.~\cite{KEKBupdate,Superctupdate}.

Besides, the investigation of a new physics beyond Standard Model (SM) in leptonic decays (such as the lepton flavor violation, CP violation and so on) the special interest represents also the study of $\tau$ decays into states containing hadrons (semileptonic decays). In such decays, in accordance with SM, the virtual state of the charged vector boson ($W^{\pm}$) is created, and then it decays into hadrons
$$\tau^{\pm}\to \nu_\tau + W^{\pm}\to hadrons.$$
This last decay we call as "hadronization of the weak charged currents".

The energy region of the last transition $(W^{\pm}\to hadrons)$ corresponds to the hadron dynamics which can not be described by the perturbative
QCD. Since the complete theory of the non-perturbative QCD is absent at present, the hadronization phenomenon, in this energy region, is described by means of weak phenomenological form factors depending on the squared hadrons invariant mass $t$ ($t$
is also the difference of the $\tau^-$ and $\nu_{\tau}$ 4-momenta
squared). Note that every separate channel with hadrons requires its own set of form factors.

The simplest semileptonic $\tau $ lepton decay is
$\tau^-\to\pi^-\nu_{\tau}.$ However, in this case, the
hadronization of the weak currents is described by the form
factors at fixed value of $t.$ There is the possibility to investigate weak form factors in the radiative transition $W\to \pi\gamma
$, where $t$, in this case, is the squared invariant mass of the $\pi  - \gamma $ system. The one-pseudoscalar meson radiative $\tau$ decay
$\tau^-\to\pi^-\nu_{\tau}\gamma$ is very suitable process to extract information about the $t-$dependence of the corresponding weak form factors.
It has been considered in the number of theoretical works \cite{K80,B86,D93,R95,G03,G10}, in which the double-differential decay rate were obtained in terms of vector $v(t)$ and axial-vector $a(t)$ form factors, as well as some distributions over the photon or pion energy and the total decay rate were estimated by
numerical integration. Certain polarization phenomena in the polarized $\tau$ decay $\tau^-\to\pi^-\nu_{\tau}\gamma$ have been considered in Ref.~\cite{R95} where the general expressions for the Stokes parameters of the emitted photon itself have been calculated. The different theoretical models of the vector and axial-vector form factors have been used in these works.

In spite of the large enough value of the integrated decay rates ratio: $R=\Gamma (\tau\to\nu_{\tau}\pi\gamma )/\Gamma (\tau\to\nu_{\tau}\pi )=1.4\cdot
10^{-2}$ (Ref. ~\cite{K80}) and  $R=1.0\cdot 10^{-2}$ (Ref.~\cite{D93}), the radiative $\tau$ decay was not recorded surely up to now.
The attempt to measure the $\tau $ radiative decays, $\tau\to l\gamma\nu\bar\nu$
($l=e, \mu$) and $\tau\to\pi\gamma\nu$, has been done using the BaBar detector of
the PEP-II asymmetric B-factory \cite{OBER13}. In this experiment, it was required that a neutral deposit of energy in calorimeter must be greater than 50 MeV.
 The BaBar experiment allowed to collect about
5$\cdot$\,10$^8$ $\tau$-pairs (that is two orders of magnitude more than it was done by CLEO \cite{CLEO2000}) but the final efficiency for the mode $\tau \to \pi\gamma\nu$ turned out very small and efficient background reduction
technique at BaBar measurements requires new investigation \cite{OBER13}. The study shows that further investigation is necessary to extract useful information about the $\tau$-lepton decay parameters.

The complete analysis of the polarized phenomena in the decay $\tau^-\to \pi^-\gamma\nu_{\tau}$ have been performed in Ref.
\cite{GKKM15} where the following polarization observables have been calculated: the asymmetry caused by the $\tau $ lepton polarization, the
Stokes parameters of the emitted photon itself and the spin correlation coefficients which
describe the influence of the $\tau $ lepton polarization on the photon Stokes
parameters. All these quantities were calculated as a functions of the photon energy
or the t variable. The azimuthal dependence of these observables  have
been calculated in Ref.\cite{GKM16}.  The so-called up-down and
right-left asymmetries are also calculated.

The amplitude of the radiative decay $\tau^-\to\nu_{\tau}\pi^-\gamma $ contains the
infrared divergence due to the inner bremsstrahlung contribution. So, the integrated
decay rates and polarization observables must depend on the photon-energy cut (or the
meson-photon invariant mass cut). Since this dependence is not trivial it is worth to
investigate it more thoroughly.

In this paper, we obtained the analytical expressions, as a function of the
photon-energy cut, for the t-distribution for the different observables for both
polarized and unpolarized $\tau$ lepton decay and illustrate results by numerical
estimations. We consider the asymmetry of the differential decay width caused by the
$\tau$ lepton polarization and the Stokes parameters of the emitted photon and the
correlation parameters, namely, the dependence of the Stokes parameters on the
polarization of the decaying $\tau$ lepton. These physical quantities are estimated
numerically for an arbitrary direction of the $\tau$ lepton polarization 3-vector in
the rest frame.

\section{Formalism}

Amplitude of the semi-leptonic radiative decay of $\tau$ lepton
\begin{equation}\label{eq:1}
\tau^-(p)\to\pi^-(q)+\gamma(k)+\nu_{\tau}(p')
%\,, \ \ \tau^-(p)\to K^-(q)+\gamma(k)+\nu_{\tau}(p')
\end{equation}
contains the structureless and structure dependent (resonance) parts
$$M_{\gamma}=Z\big(M_{IB}+M_R\big)\,, \ \ Z=eG_FV_{ud}F_{\pi}\,.$$
Here
$e^2/4\pi=\alpha=1/137\,,$ $G_F=1\,.166\cdot 10^{-5}GeV^{-2}$ is
the Fermi constant of the weak interactions,
%. For the pion in final state $V_P=V_{ud}=0.9742 \ (V_P=V_{us}=0.2253) $ is
$V_{ud}=0.9742$ is the corresponding element of the CKM-matrix,
% \cite{CKM},
$F_{\pi}=92.42 $MeV
% \ (F_P=F_{K}=113 MeV )$
is the constant which determines the decay
$\pi^-\rightarrow\mu^-\bar\nu_{\mu}.$
% \ (K\,^-\rightarrow\mu^-\bar\nu_{\mu}).$

The structureless amplitude is responsible for the radiation by $\tau$ lepton and pseudoscalar
meson within the point-like scalar electrodynamics

\begin{equation}\label{eq:2}
iM_{IB}=M\bar u(p')(1+\gamma_5)\Big[\frac{\hat
k\gamma^{\mu}}{2(kp)}+\frac{Ne_1^{\mu}}{(kp)(kq)}\Big]u(p)\varepsilon^*_{\mu}(k)\,,
\end{equation}
where M is the $\tau$ lepton mass, and
$\varepsilon_{\mu}(k)$ is the photon polarization 4-vector.
The rest notation is
$$e_1^{\mu}=\frac{1}{N}\big[(pk)q^{\mu}-(qk)p^{\mu}\big]\ , \ \
(e_1k)=0\ , \ e_1^2=-1\ , $$
$$ N^2=2(qp)(pk)(qk)-M^2(qk)^2-m^2(pk)^2\ ,$$
where $m$ is the $\pi-$meson mass.

In the $\tau-$lepton rest frame the differential width reads
\begin{equation}\label{eq:3}
d\Gamma=\frac{1}{4M(2\pi)^5}|M_{\gamma}|^2\, d\,\Phi\,, \ d\,\Phi=\frac{d^3k}{2\omega}\frac{d^3q}{2\epsilon}\delta(p'^{2})\,,
\end{equation}
where $\omega\, (\epsilon)$ is the energy of the photon (pseudoscalar meson).

The structure dependent amplitude allows for the possibility of the virtual radiative transition $W\,^-\to \pi^-+\gamma.$
It is expressed in terms of complex vector $v(t)\,, \ t=(q+k)^2,$ and axial-vector $a(t)$ form factors
which describe, on the phenomenological level, the hadronization process of the charged weak currents \cite{K80,B86,D93,R95,G03,G10,R14,GKKM15,GKM16}

\begin{equation}\label{eq:4}
iM_R=\frac{Z}{M^2}\bar
u(p')(1+\gamma_5)\Big\{i\gamma_{\alpha}(\alpha\mu
kq)v(t)-\big[\gamma^{\mu}(qk)-q^{\mu}\hat k
\big]a(t)\Big\}u(p)\varepsilon^*_{\mu}(k)\ ,
\end{equation}
$$(\alpha\mu kq) = \epsilon^{\alpha\mu\nu\rho}k_{\nu}q_{\rho}\ ,
\ \epsilon^{0123}=+1\ , \ \
\gamma_5=i\gamma^0\gamma^1\gamma^2\gamma^3\,, \
Tr\gamma_5\gamma^{\mu}\gamma^{\nu}\gamma^{\rho}\gamma^{\lambda}=-4i\epsilon^{\mu\nu\rho\lambda}\,. $$

Our normalization of the structural amplitude is, up to overall phase factor, the same as in Ref.~\cite{R95} where the authors use as the value
of the pion decay constant the quantity $\sqrt{2} F_{\pi}$ as compared with our's one but they compensate this distinction by introducing the factor $\sqrt{2}$
in the denominator of the value $Z$ in Eqs.~(\ref{eq:3}\,,\ref{eq:4}). As concerns some other cited above papers, the amplitudes of Refs~\cite{D93,G03}
can be derived from our ones by the substitutions
$$v(t)\to \frac{M^2\,F_V(t)}{\sqrt{2}\,m\,F_P}\,, \ \ a(t)\to -\frac{M^2\,F_A(t)}{\sqrt{2}\,m\,F_P}\,,$$
and the amplitudes of the Ref.~\cite{G10, R14} by
$$v(t) \to -\frac{M^2\,F_V(t)}{F_P}\,, \ \ a(t) \to \frac{2 M^2\,F_A(t)}{F_P}\,.$$

The main interest to study decay (\ref{eq:1}) is the extraction of information about these form factors.
As we note in the Introduction, in
the high luminosity electron-positron colliders, such as SuperKEKB (Japan) and Super $c-\tau$ (Russia) \cite{KEKBupdate,Superctupdate}, it is expected to obtain about
10$^{10}$ events with polarized $\tau^+\tau^-$ pairs. It means that, in accordance with the theoretical estimation, at these facilities one can receive and analyse more than 10$^6$ radiative decays with pion only.

If $\tau$ lepton, in decay (\ref{eq:1}), is unpolarized then the decay width depends only on two variables which can be chosen as energies of the photon and pseudoscalar meson, but to study form factors it is convenient to chose the photon energy and variable $t,$ which vary in the limits
$$ t_-~\leq ~ \omega~ \leq ~ t_+\,, \ t_- =\frac{t-m^2}{2M}\,, \ t_+ =\frac{M(t-m^2)}{2t}\,; \ \ m^2 ~ \leq ~ t ~ \leq ~ M^2\,.$$
In a number of the cited papers, the  authors perform the analytical integration over the photon energy inside the above limits and write down the expression for $d\,\Gamma(t)/d\,t.$

But because of the infrared behaviour of the structureless contribution
that exhibits itself by convergence of this contribution at $t\to m^2,$
the low-energy radiation leads to large background
in studying the form factors. That is why it is need to cut the low-energy photon radiation and to select only the events with the restriction $\omega>X.$
Some optimal value of $X$ has to be chosen in order to reduce significantly the structureless contribution and do not lose the resonance one. In this case, the
region of the variables $\omega$ and $t$ is shown in Fig.~1 (right panel), where
$$t_{x2}=\frac{m^2 M}{M-2 X}\,, \ t_{x1}=2 M X -m^2\,.$$
We see that in  the region {\bf 2}, where $t>t_{x2},$ the $t-$distribution remains unchanged, whereas in the region {\bf 1} $(t_{x1}<t<t_{x2})$ it
depends on the value of the photon-energy cut and requires some modification. Further, we perform the corresponding calculations, give the
modified $t-$distributions for different observables for both polarized and unpolarized $\tau$ lepton decays and illustrate results by numerical estimations.

\begin{figure}

\includegraphics[width=0.40\textwidth]{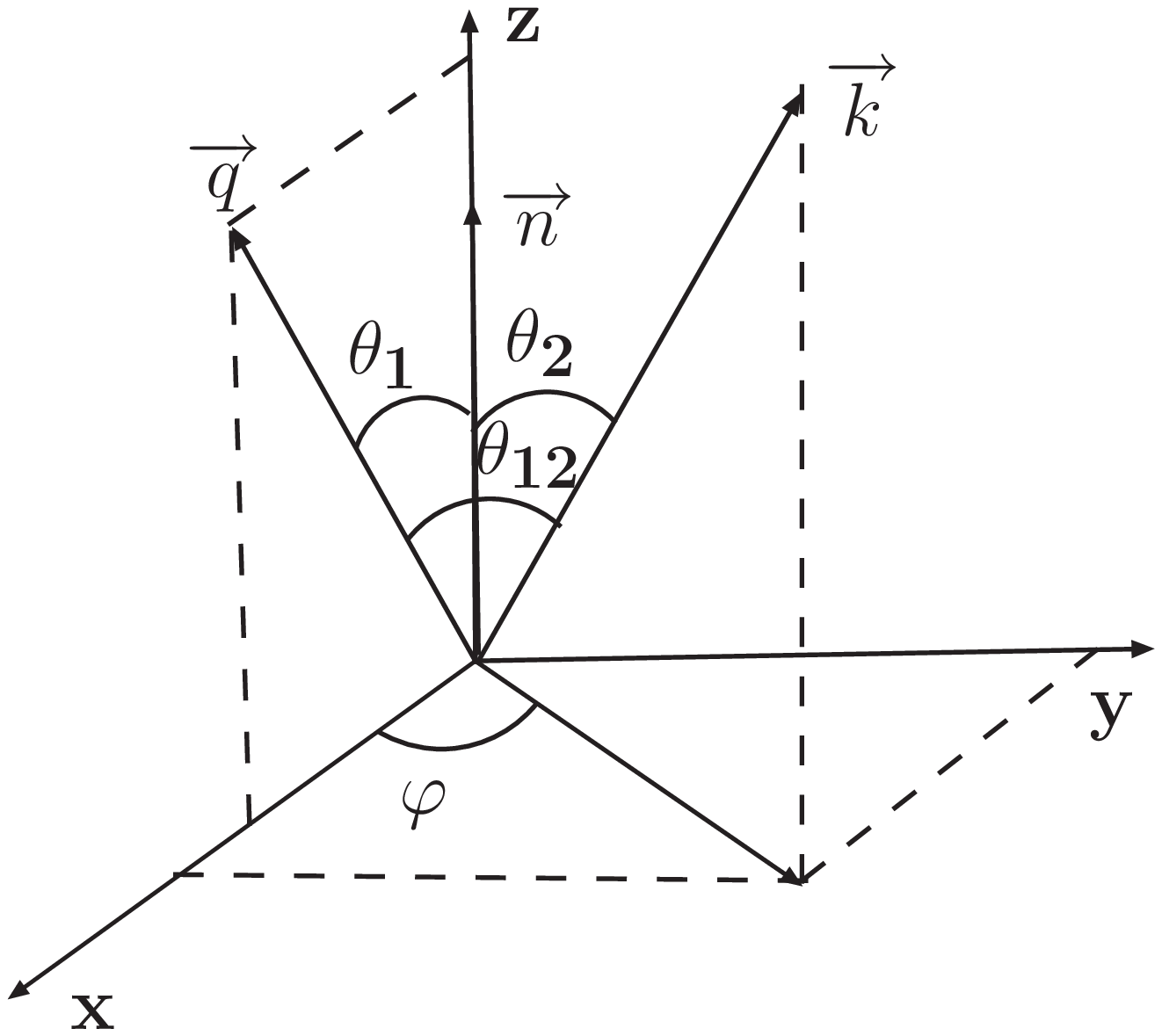}
%\hspace{0.4cm}
\includegraphics[width=0.32\textwidth]{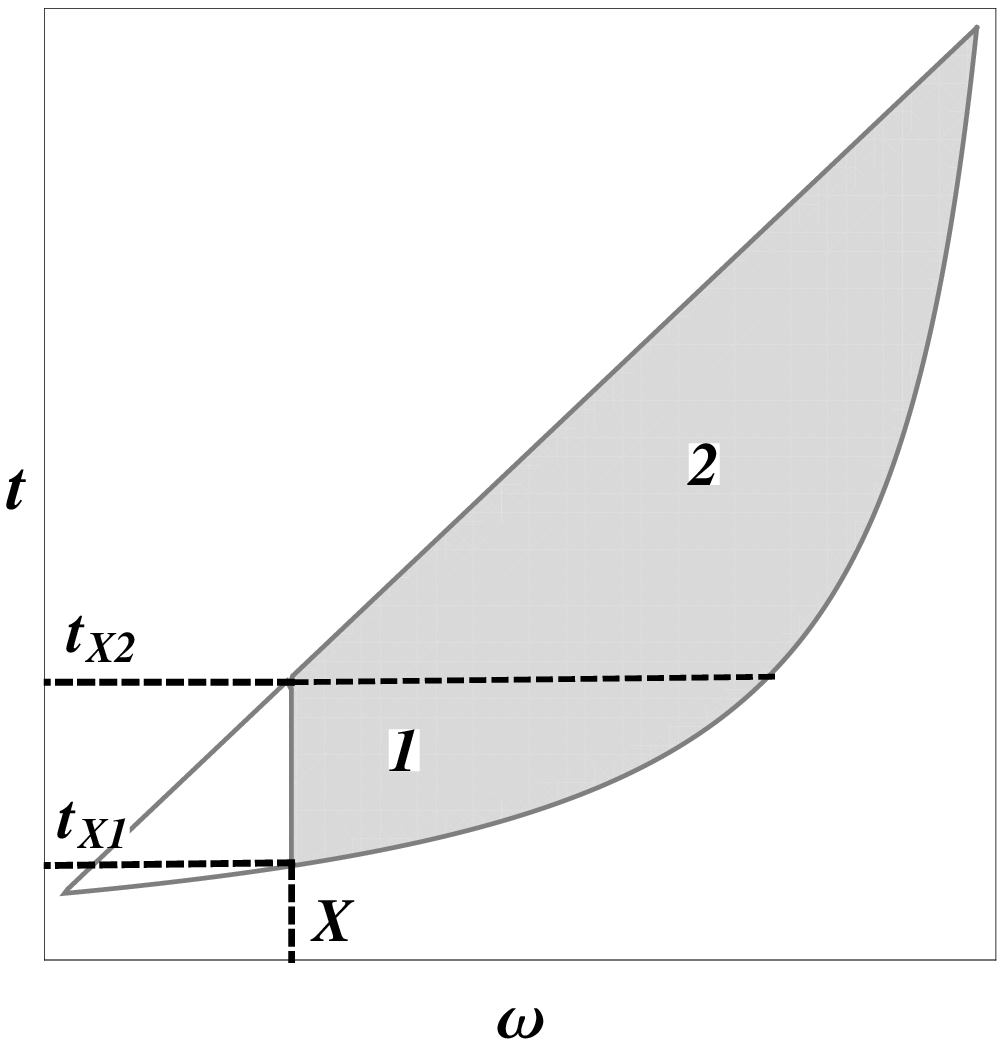}

\caption{Definition of the angles for a
 polarized radiative $\tau$ decay at rest (left panel)\,; the region of variation of the photon energy $\omega$ and
 variable $t$ for events with $\omega >X$(right panel).}
\end{figure}

In general case, to study polarization effects, it is convenient to represent the matrix element squared in the following form
\begin{equation}\label{eq:5}
|M_\gamma|^2=\Sigma + \Sigma_i,\, \ \Sigma=T^{\mu\nu}(e_{1\mu}e_{1\nu}+e_{2\mu}e_{2\nu})\,, \ \Sigma_1=T^{\mu\nu}(e_{1\mu}e_{2\nu}+e_{2\mu}e_{1\nu})\,,
\end{equation}
$$\Sigma_2=-iT^{\mu\nu}(e_{1\mu}e_{2\nu}-e_{2\mu}e_{1\nu})\,, \ \ \Sigma_3=T^{\mu\nu}(e_{1\mu}e_{1\nu}-e_{2\mu}e_{2\nu})\,, \ e_2^\mu=\frac{(\mu pqk)}{N}\,,$$
where the expression for the tensor $T^{\mu\nu}$ is given in \cite{GKKM15}.
If $\tau$ lepton is polarized then
$$T_{\mu\nu}=T^0_{\mu\nu}+T^{^S}_{\mu\nu}\,, \ \Sigma=\Sigma^0+\Sigma^{^S}\,, \ \Sigma_i=\Sigma_i^0+\Sigma_i^{^S}\,,$$
where the upper index "$0 ~(S)$" corresponds to unpolarized (polarized) $\tau$ lepton, and we can define the polarization asymmetry $A^{^S}$, the Stokes parameters of the photon itself $\xi_i$ and the correlation parameters $\xi_i^{^S}$ as
\begin{equation}\label{eq:6}
A^{^S}=\frac{\Sigma^{^S}d\,\Phi}{\Sigma^{^0}d\,\Phi}\,, \ \
\xi_i=\frac{\Sigma^{^0}_id\,\Phi}{\Sigma^{^0}d\,\Phi}\,, \ \
\xi^{^S}_i=\frac{\Sigma^{^S}_id\,\Phi}{\Sigma^{^0}d\,\Phi}\,.
\end{equation}

Thus, to analyze the polarization phenomena in the process (1), we
have to study both the spin-independent and spin-dependent parts
of the differential width. In accordance with Eq.~(\ref{eq:3}), they
are
$$\frac{d\,\Gamma_0}{d\,\Phi}=g\Sigma^{^0}, \ \ \frac{d\,\Gamma^{^S}_0}{d\,\Phi}=g\Sigma^{^S},
\ \frac{d\,\Gamma_i}{d\,\Phi}=g\Sigma^{^0}_i, \
\frac{d\,\Gamma^{^S}_i}{d\,\Phi}=g\Sigma^{^S}_i, \
g=\frac{1}{4\,M(2\,\pi)^5}\,.$$ We note that by partial integration
in the numerators and denominator in the relations (\ref{eq:6}), we can define
and study also the corresponding reduced polarization parameters. It is obvious that the quantities $A^{^S}$ and $\xi^{^S}_i$ vanish if the
full angular integration is performed. Therefore, to study corresponding polarization effects we have to extract the
contributions to the differential decay width from certain regions of the photon angular phase space.

\section{Angular phase space and classification of the polarization-dependent observables}

As we mentioned above, the decay width does not depend on any angles if $\tau$ lepton is unpolarized. In this case, we can define the
differential decay width and the Stokes parameters of a photon \cite{R95, GKKM15, GKM16}. But there is additional 3-vector in the rest frame
of the polarized $\tau$ which is just its polarization $S^\mu=(0, {\bf n}), {\bf n}^2=1$. Therefore, the angular dependence of the decay width arises
due to appearance of the polarization dependent terms containing $(Sk), \ (Sq)$ and $(Spqk)=\epsilon_{\mu\nu\lambda\rho} S^\mu p^\nu q^\lambda k^\rho.$
In the $\tau$ lepton rest frame we chose the coordinate system with $Z$ axis along direction ${\bf n} $ and 3-vector ${\bf q}$ in the plane $(Z\,X)$ as it is shown in Fig.~1 (left panel) and have
\begin{equation}\label{eq:7}
(Sq)= -|{\bf q}|c_1\,, \ (Sk)=-\omega c_2\,, \ (Spqk)=M|{\bf q}|\omega s_1\,s_2 \sin{\phi}\,, \ s_1\,s_2 \sin{\phi}=sign(\sin{\phi}) K\,,
\end{equation}
$$K=\sqrt{1-c_1^2-c_2^2-c_{12}^2+2c_1 c_2 c_{12}}\,,$$
where we use the notation $c_{i}=\cos{\theta_{i}}\,, \ s_{i}=\sin{\theta_{i}}\,,$ (for the definition of the angles see Fig.~1).

The phase space in Eq.~(\ref{eq:3}) can be written in terms of the photon energy $\omega$, variable $t$ and the angles defined in Fig.~1 as in Ref.~\cite{GKM16}
\begin{equation}\label{eq:8}
d \Phi = \frac{2\pi\,d\omega\,dt}{16\,M} \delta(c_{12}-c_1 c_2-s_1 s_2 \cos{\phi})dc_1 dc_2 d\phi\,,
\end{equation}
where the factor $2 \pi$ reflects the freedom in choosing of the $(Z\,X)$ plane.
In this paper, we do not consider the azimuthal distributions, nevertheless we will
distinguish the events with the full azimuthal integration $(0<\phi<2\pi),$ and with the separate one in the regions $(0<\phi<\pi)$ and $(\pi<\phi<2\pi).$
In the first case, only the terms containing $(Sq)$ and $(Sk)$ will contribute, whereas in the second one we have the possibility to study the effects
due to the term $(Spqk)$ taking the difference of events in these regions.

For the full azimuthal integration one derives
\begin{equation}\label{eq:9}
\delta(c_{12}-c_1 c_2-s_1 s_2 \cos{\phi})dc_1 dc_2 d\phi=\frac{2 dc_1 dc_2}{K}\,, \ c_{1-}< c_1 < c_{1+}\,, \ c_{1\pm}=c_2 c_{12} \pm s_2 s_{12}\,.
\end{equation}
We will analyse events with the pseudoscalar meson in the whole angular phase space, thus the integration over $c_1$ has to be performed
from $c_{1-}$ up to $c_{1+}.$ The necessary integrals are
$$\int\frac{d c_1}{K}=\pi\,, \ \ \int\frac{c_1 d c_1}{K}=\pi c_2 c_{12}\,,$$
and the quantity $|{\bf q}|\,c_{12}$, which arises after such integration of the terms containing $(Sq),$ can be expressed in terms of the photon energy and
variable $t,$ namely
\begin{equation}\label{eq:10}
|{\bf q}|\,c_{12}=\frac{M^2+t}{2\,M}-\frac{t-m^2}{2\,\omega}-\omega\,.
\end{equation}

If $\tau$ lepton is unpolarized, the differential decay width does not depend on any angles and the event number in the upper hemisphere ($0<c_2$)
equals to the event number in the lower one ($0>c_2$). In the case of polarized $\tau$ this equality is broken and we can define the corresponding asymmetry
as a ratio of the difference and sum of the events number accumulated in the upper and lower hemispheres. Such kind of asymmetry we call as the {\it up$-$down} one. In principle, the same situation is valid if we want to analyse the influence of $\tau$-lepton polarization on the photon Stokes parameters.

We can also separate events with the photon in the right $(0<\phi<\pi\,, \ \sin{\phi}>0)$ and left $(\pi<\phi<2\pi\,, \ \ \sin{\phi}<0)$ hemispheres for all its polar angles. The ratio of the difference and sum of the events number accumulated in the right and left hemispheres we call as the {\it right$-$left} asymmetry. This asymmetry appears due to the factor $(Spqk)$ in the matrix element squared, and integration of this factor over the polar angles in the right hemisphere reads
\begin{equation}\label{eq:11}
\int\limits_{-1}^1 d\,c_2 \int\limits_{c_{1-}}^{c_{1+}}d\,c_1 \frac{(Spqk)}{K}=\pi\,M\,|{\bf q}|\omega s_{12}\,, \ |{\bf q}|\omega s_{12}=
\sqrt{t(\omega-t_-)(t_+-\omega)}\,,
\end{equation}
whereas in the left hemisphere this factor contributes with the opposite sign.

\section{Distribution over the invariant mass squared of the photon-pseudoscalar meson system}

Because the polarization observables are normalized by the unpolarized decay width (see relations (\ref{eq:6})), we have firstly to write
the analytic $t-$distribution for the unpolarized case taking into account the cut on the photon energy $\omega>X$. In the region {\bf \large{2}}, this distribution is standard (as in the above cited papers) and only contribution of the region {\bf \large{1}} requires some modification. The integration of the
double differential decay width over the photon energy in this region results

\begin{equation}\label{eq:12}
\frac{d\Gamma_{0x}}{d\,t}=P\big[I_{0x}(t,X)+(|a(t)|^2+|v(t)|^2)A_{0x}(t,X)+Re[a^*(t)v(t)]D_{0x}(t,X)
\end{equation}
\[+Re(a(t))B_{0x}(t,X)+Re(v(t))C_{0x}(t,X)\big]\,, \ \ P=\frac{Z^2}{2^8\pi^3M^2}\,,\]
\[I_{0x}(t,X)=\frac{4 M L_x \left[m^4-2 M^2 \left(-m^2+M^2+t\right)+t^2\right]}{m^2-t}\]
\[+\frac{4 M \Omega _x \left[M X \left(t \left(2
   M^2-t\right)-m^4\right)+t \left(m^2-M^2\right) \left(m^2-t\right)\right]}{t X \left(m^2-t\right)^2}, \ L_x=\ln{\frac{M(t-m^2)}{2X t}}\,,\]
\[A_{0x}(t,X)=-\frac{\Omega_x \left[\left(m^2-t\right)^2 \left(M^4-3 t^2\right)+2 M t X \left(\left(t-m^2\right) \left(M^2+3 t\right)-4
   M t X\right)\right]}{3 M^4 t^2}\,, \]
\[B_{0x}(t,X)=\frac{4 L_x \left(t-m^2\right) \left(m^2-2 M^2-t\right)}{M}-\frac{4 \Omega_x \left[\left(m^2-t\right) \left(-m^2+M^2+3
   t\right)+2 M t X\right]}{t \left(m^2-t\right)}\,,\]
\[C_{0x}(t,X)=\frac{4 L_x\left(m^2-t\right)^2}{M}+\frac{4 \left(t-m^2\right) \Omega_x}{t}\,,\]
\[D_{0x}(t,X)=-\frac{2 \Omega_x\left(m^2-t\right) \left(m^2+2 M X-t\right)}{M^4}\,, \ \ \Omega_x=2t(X-\omega_+)\,.\]

The resonance contribution containing $Re[a^*(t)v(t)]$ is absent for the entire photon phase space. It is easy to verify this noting that
$D_{0x}(t,\omega_{min})=0$ because the factor $(m^2+2 M X-t)$ is just $2M(X-t_-)\,.$
The effect of the photon-energy cut, for the unpolarized case is shown in Fig.~2.

The numerical calculations of the analytical expressions for various observables were performed with two sets of the parameters which are taken from the chiral effective theory with resonances needs for the estimation of the form factors.
 The first set is chosen as given in Ref. \cite{GKKM15} and called there (and in this paper) as the set 1. The second one is taken from Ref. \cite{DRPP10} and in the calculations we call it as the set 3.  It was obtained using
a good fit for the pions invariant-mass squared distribution of the decay width in $\tau\to\pi\pi\pi \nu_\tau,$ measured by ALEPH \cite{ALEPH98} . The essential point of the set 3, as compared with the set 1, is the difference in accounting of the $\rho'$ resonance to the vector form factor. For the set 1 we use (see for details \cite{GKKM15})
\begin{equation}\label{eq:v}
v(t)=-f_V(t)\frac{M^2}{m F_\pi\sqrt{2}}\,, \ f_V(t)= \frac{\sqrt{2}~m}{F_\pi}\Big[\frac{N_c}{24 \pi^2}+\frac{4\sqrt{2}~h_V F_V~t}{3~m_\rho [m_\rho^2-t-i m_\rho \Gamma_\rho(t)]}\Big]
\end{equation}
and
\[F_A=0.1368~GeV, \ F_V = 0.1564~ GeV, \ m_\rho=0.7755~GeV. \]
For the set 3, in accordance with Eqs.~(32,~33) of the Ref.~\cite{DRPP10}~, we substitude in (\ref{eq:v})
\[\frac{1}{m_\rho^2-t-i m_\rho \Gamma_\rho(t)} \rightarrow \frac{1}{1+\beta_{\rho'}}\Big[\frac{1}{m_\rho^2-t-i m_\rho \Gamma_\rho(t)}+
\frac{\beta_{\rho'}}{m_{\rho'}^2-t-i m_{\rho'} \Gamma_\rho'(t)}\Big],\]
 where the $\rho'$ width is given by its two pion decay
\[\Gamma_{\rho'}(t)=\Gamma_{\rho'}(m_{\rho'}^2)\frac{m_{\rho'}}{\sqrt{t}}\Big(\frac{t-4 m^2}{m_{\rho'}^2-4 m^2}\Big)^{3/2}\,.\]
The corresponding values of the model parameters are
\[F_V=0.18~ GeV, \ F_A=0.149~ GeV, \ m_\rho=0.775~ GeV, \ m_{\rho'}=1,485~ GeV, \]
 \[\beta_{\rho'}=-0.25,\ \Gamma_{{\rho'}}(m_{\rho'}^2)=400~ MeV\,.\]

As concerns the axial-vector form factor, for the set 1
\begin{equation}\label{eq:a}
a(t)=-f_A(t)\frac{M^2}{m F_\pi\sqrt{2}}\,, \ f_A(t)=\frac{\sqrt{2}~m}{F_\pi}\Big[\frac{F_A^2}{m_{a_1}^2-t-i m_{a_1} \Gamma_{a_1}(t)}
+\frac{F_V(2 G_V-F_V)}{m_\rho^2}\Big]
\end{equation}
with $m_{a_1}=1.230~GeV, \  G_V=0.06514~ GeV\,$ and $\Gamma_{a_1}(t)$ as given by Eq.~(49) in \cite{GKKM15} (it taken from \cite{KS90, EU03}).
For the set 3 we use the same form of $a(t)$ but with $m_{a_1}=1.120, \ G_V=(F_V^2-F_A^2)/F_A$ and $\Gamma_{a_1}(t)$ in accordance with Eq.~(17) in
 Ref.~\cite{NPRS13}.
As concerns the  contribution of the $a_1'$ into axial-vector form factor, we do not account it keeping in mind the result of the  Ref.~\cite{AV15} where it was shown that the description of the decay $\tau^-\to\pi^-\nu_{\tau},$ it is enough to use standard NJL model, where only the axial-vector meson $a_1$ was taken into account.

\begin{figure}

\includegraphics[width=0.30\textwidth]{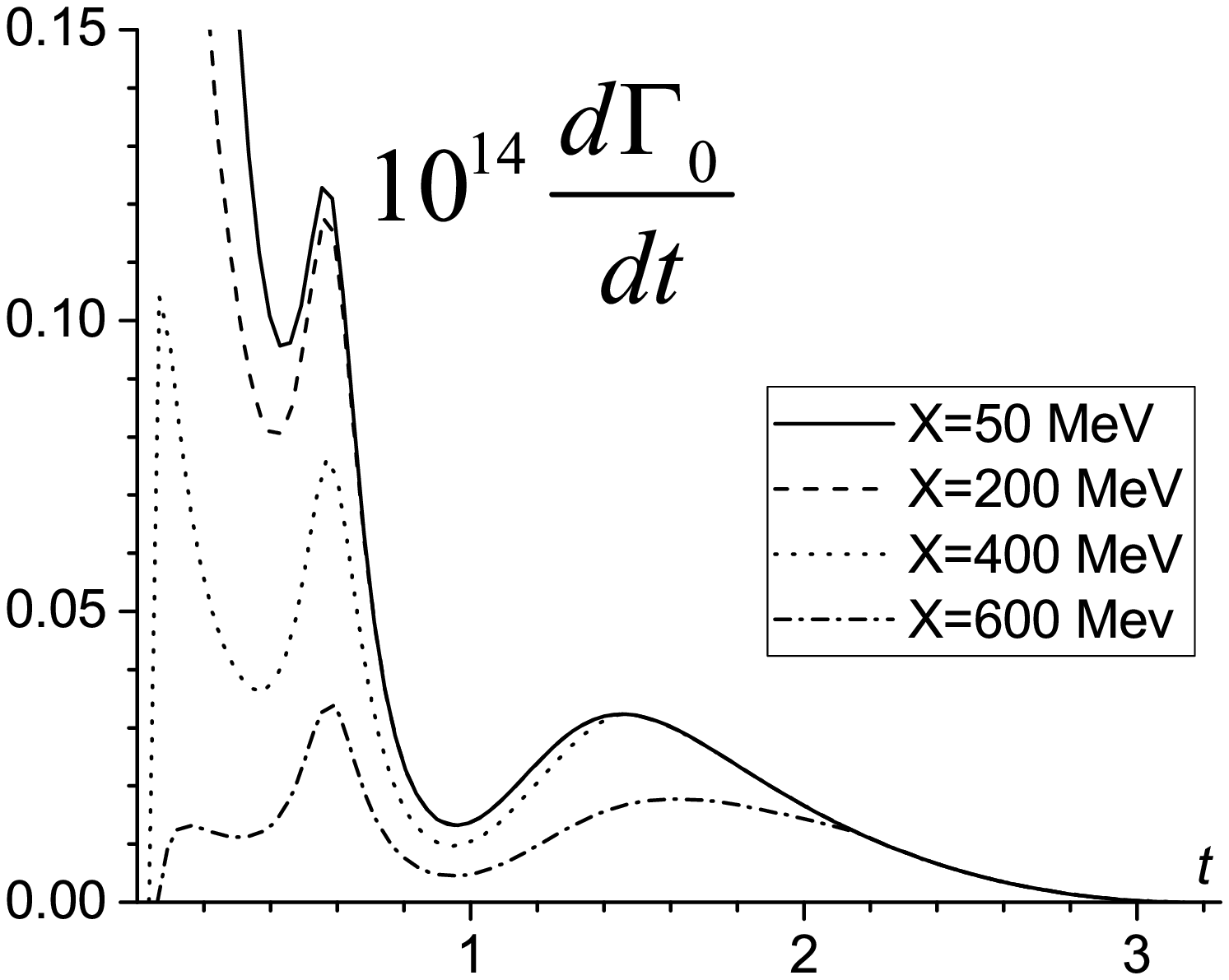}
\includegraphics[width=0.30\textwidth]{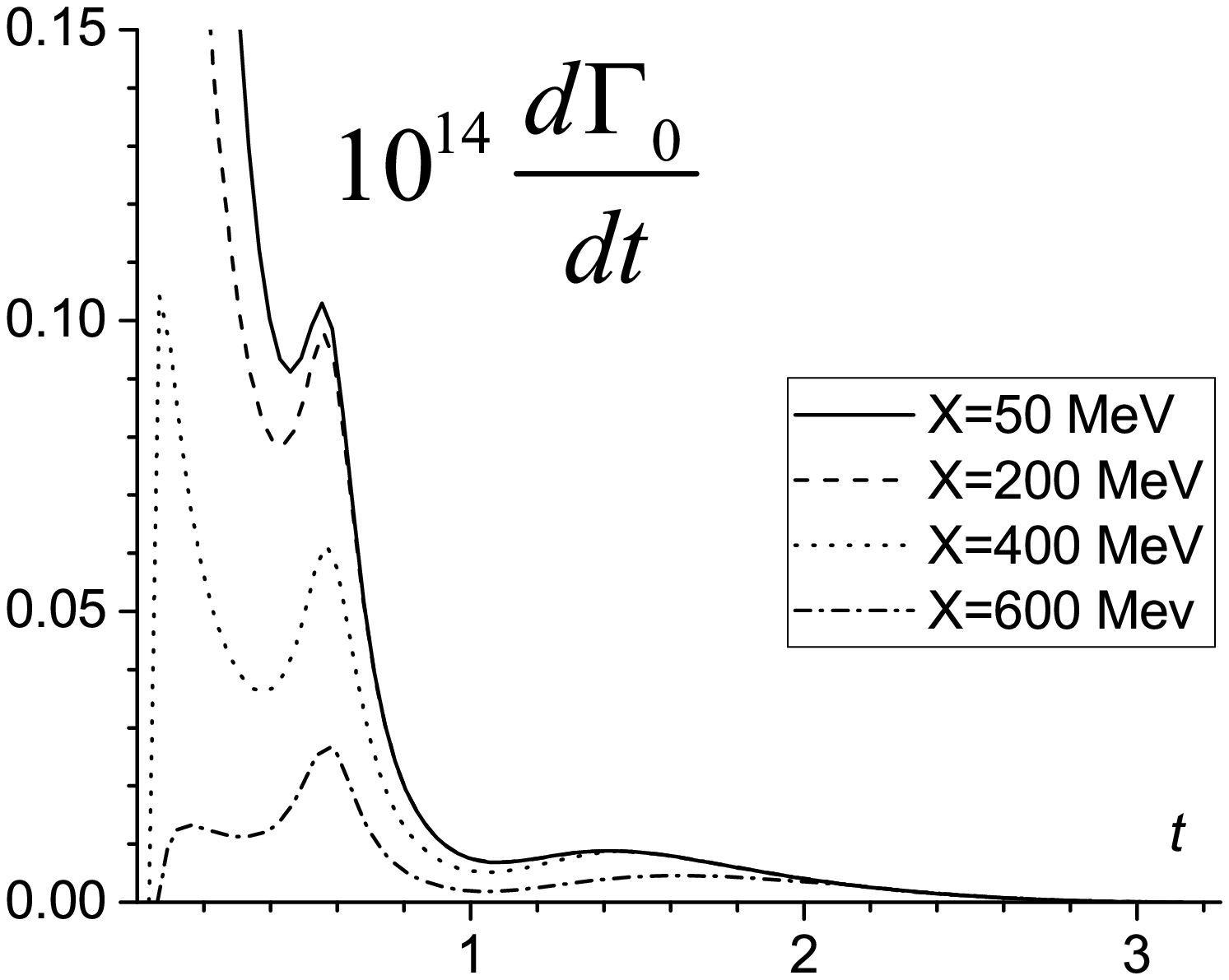}
\includegraphics[width=0.30\textwidth]{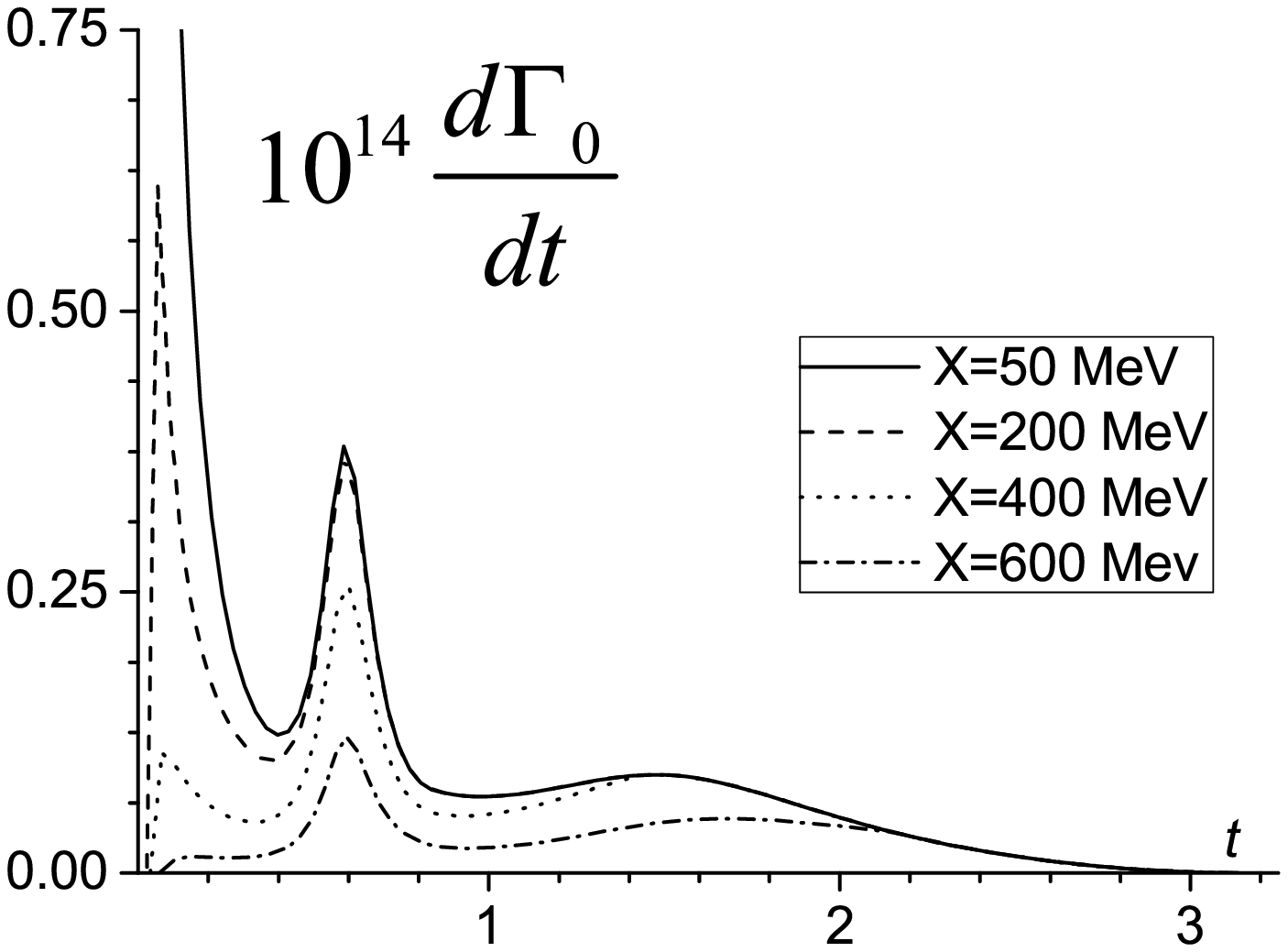}
\vspace{0.5cm}

\includegraphics[width=0.30\textwidth]{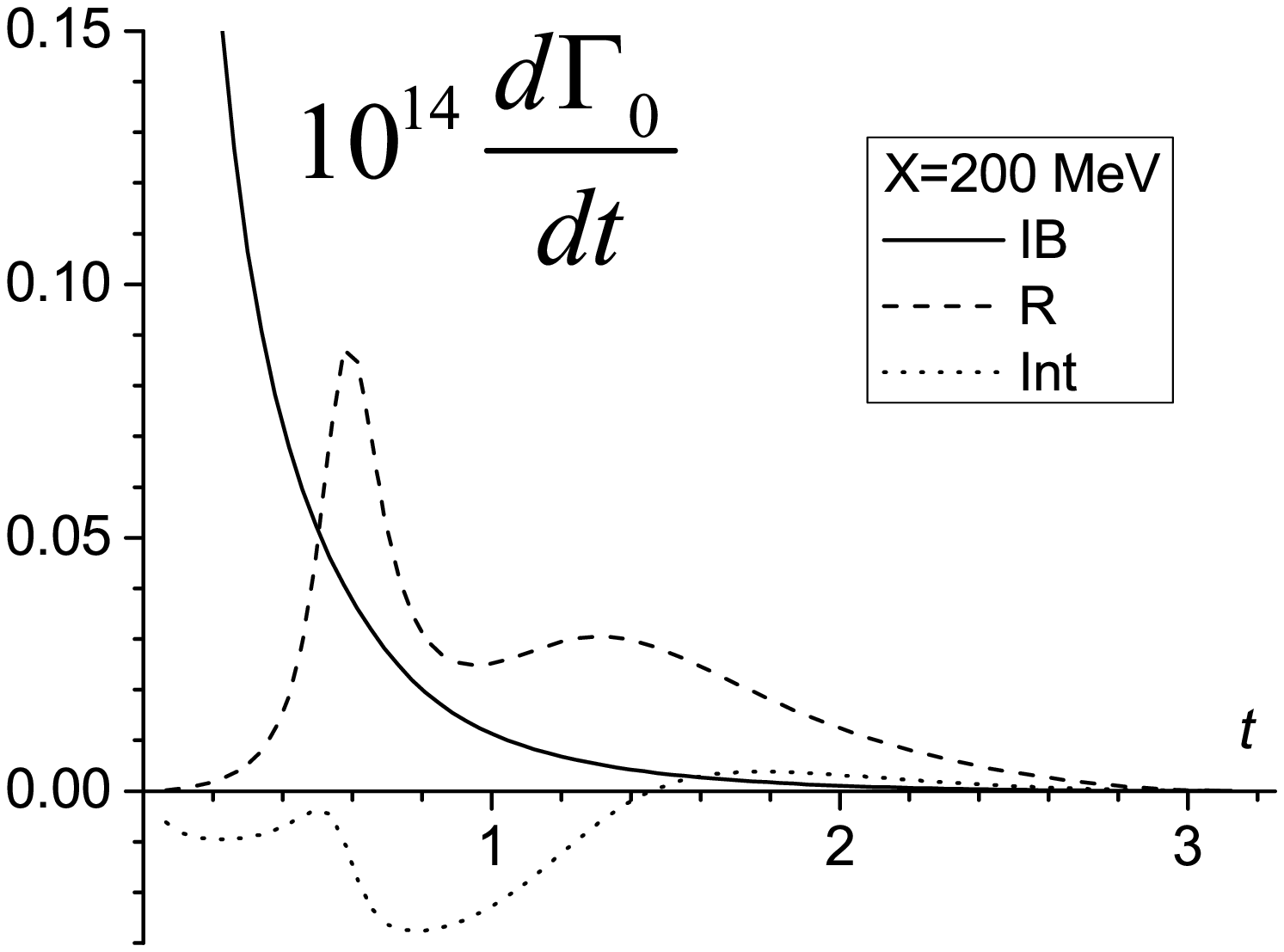}
\includegraphics[width=0.30\textwidth]{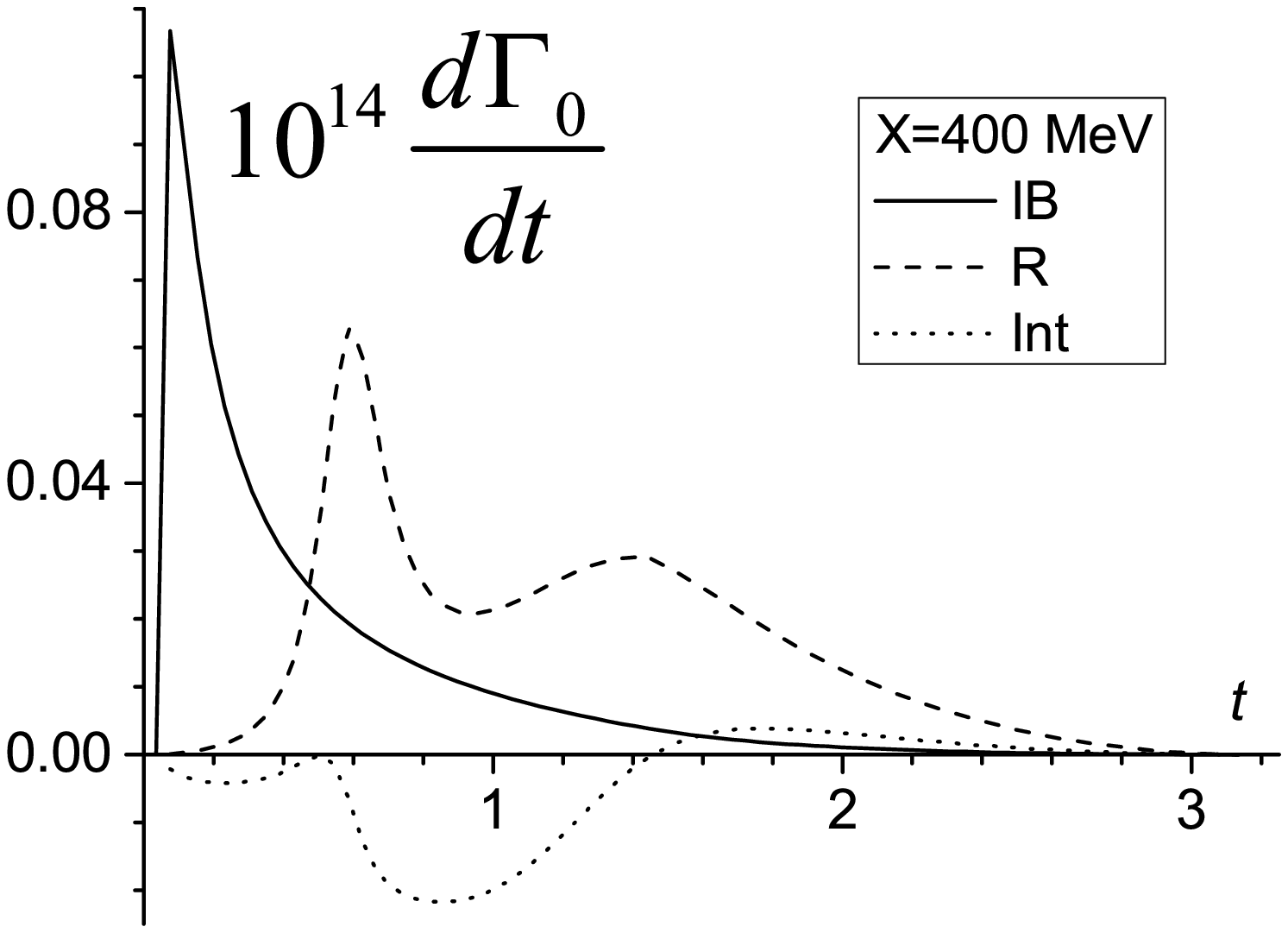}
\includegraphics[width=0.30\textwidth]{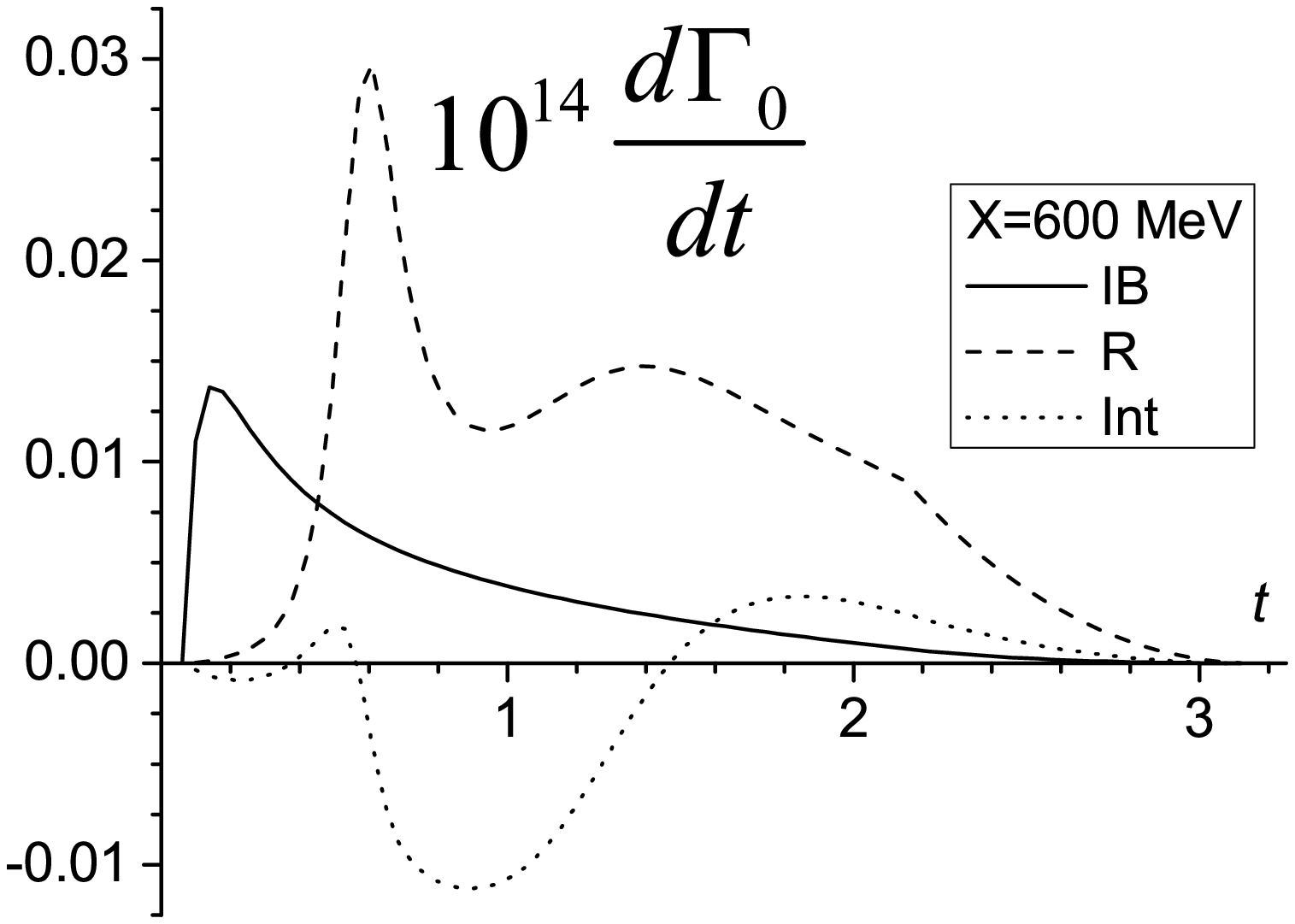}
\vspace{0.5cm}

\includegraphics[width=0.30\textwidth]{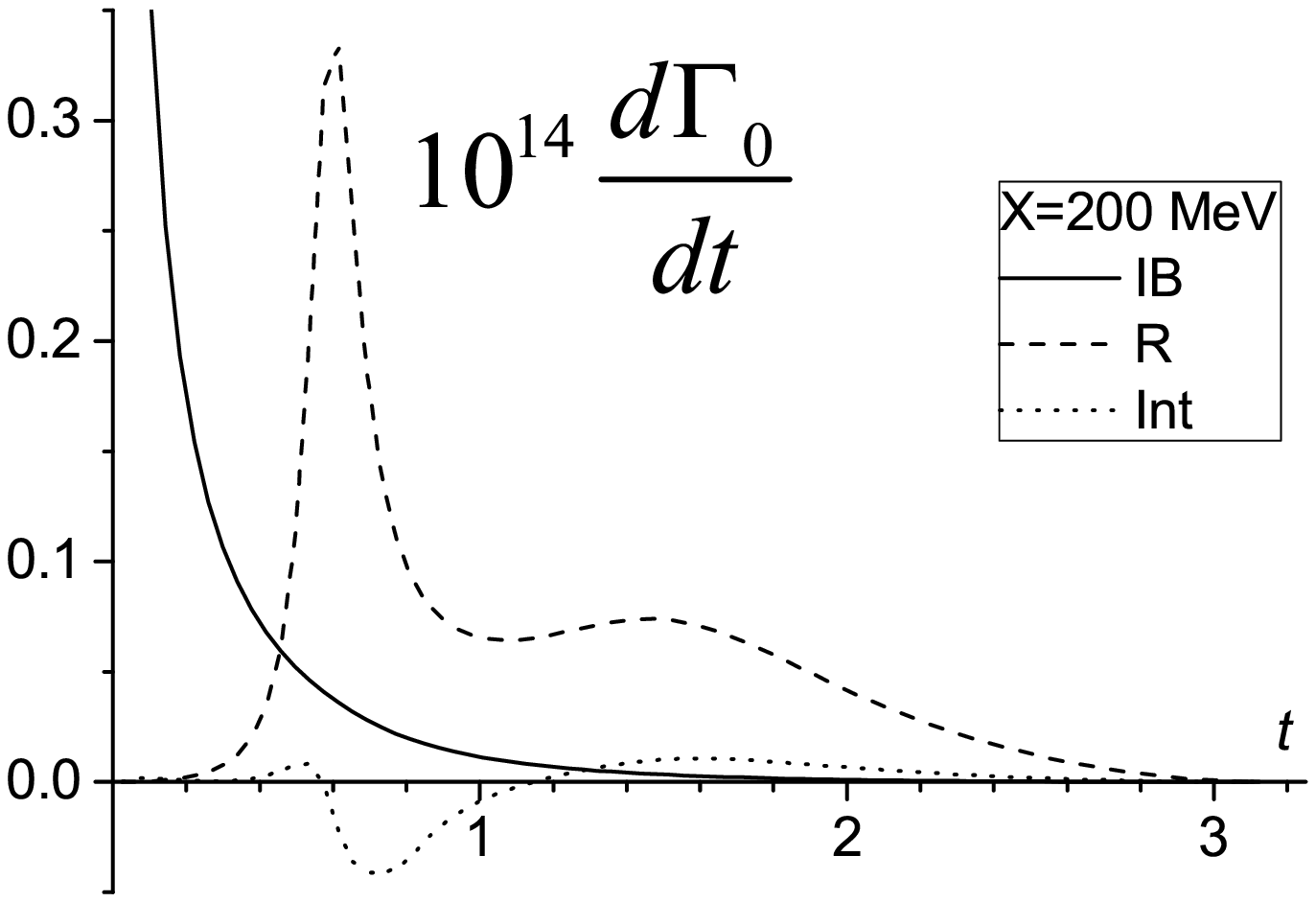}
\includegraphics[width=0.30\textwidth]{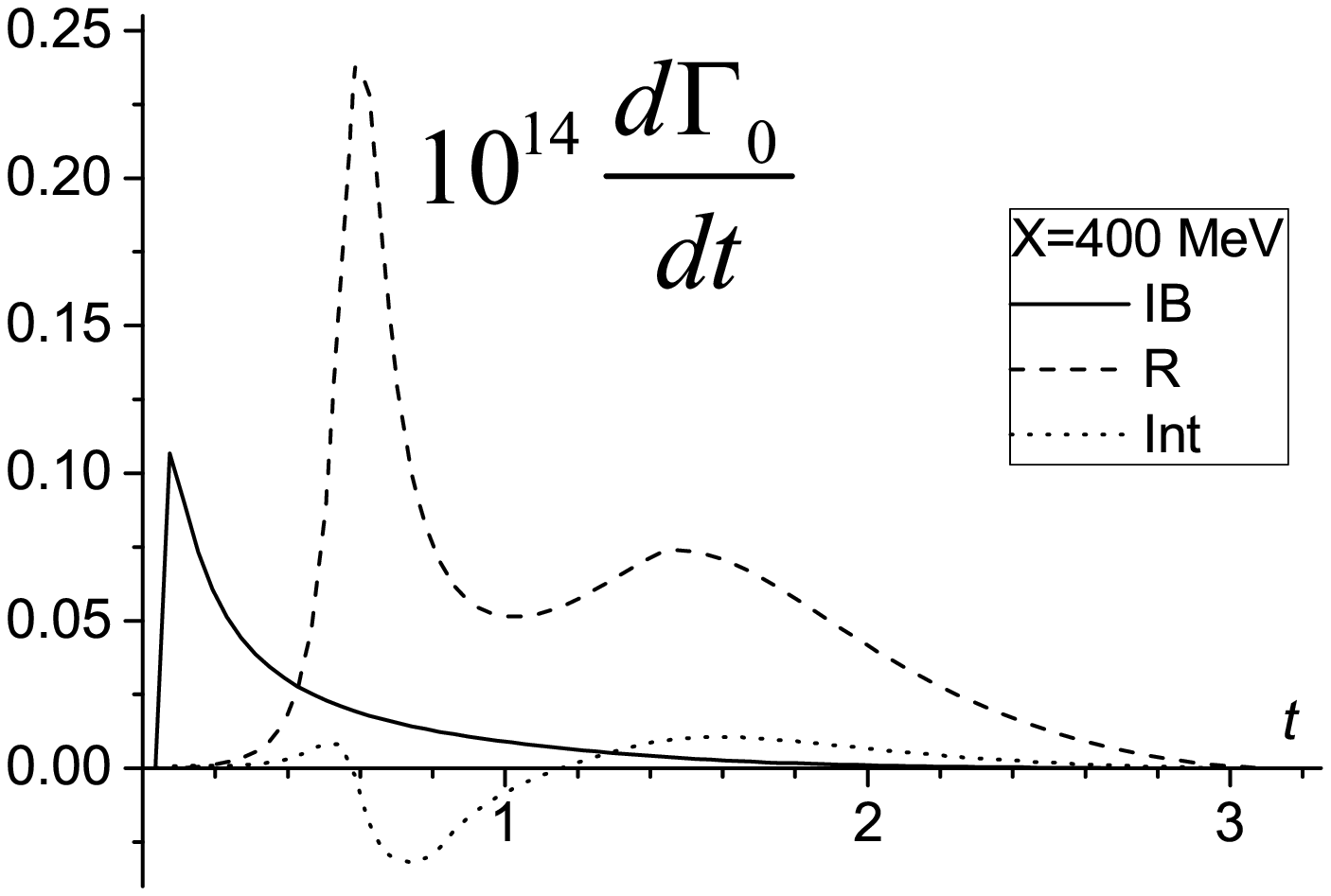}
\includegraphics[width=0.30\textwidth]{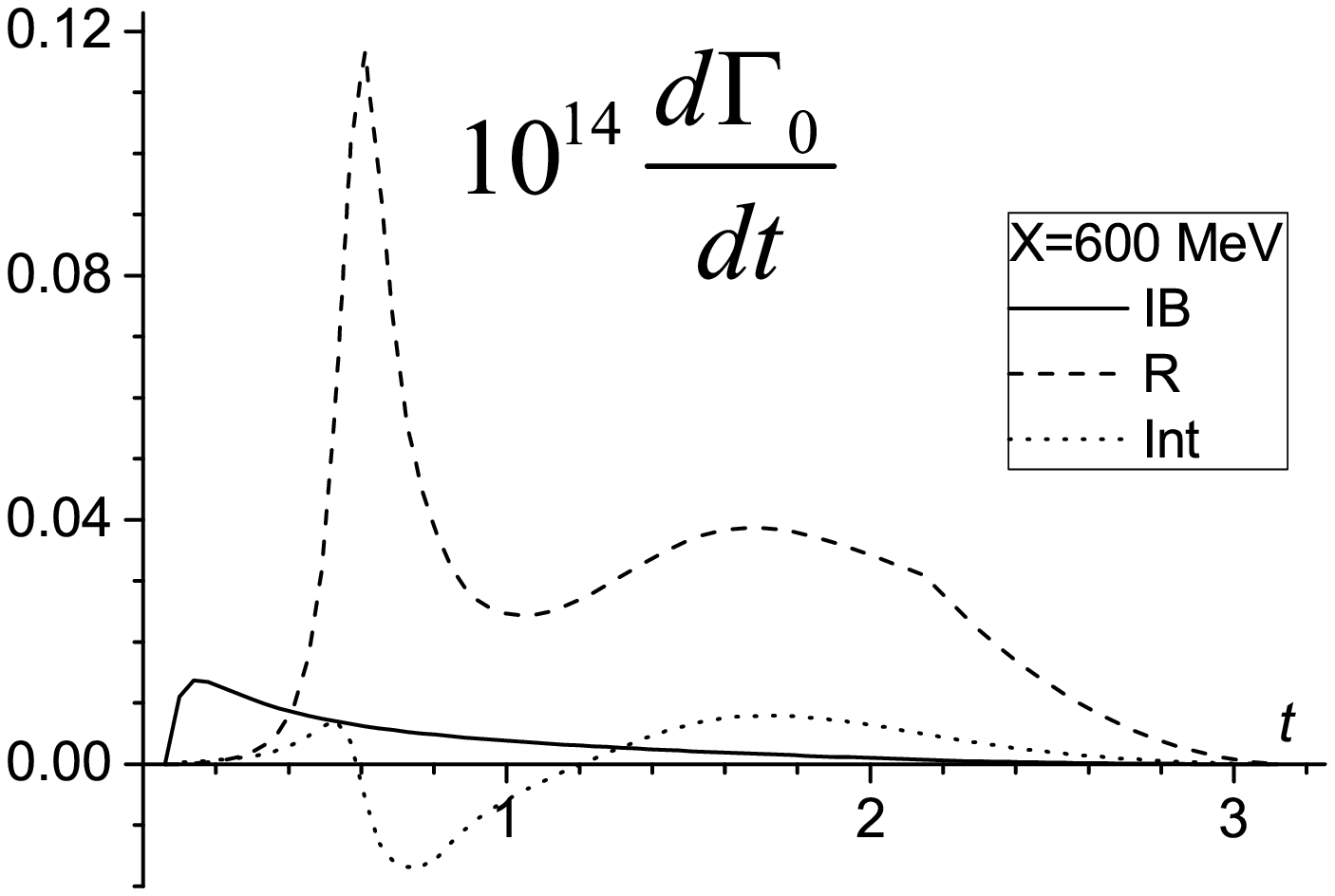}

\caption{The $t-$distribution of the radiative decay width of the unpolarized $\tau$ lepton, in GeV$^{-1}$
and $t$ in GeV$^2$ . In the region {\bf {\large 2}} we used the
analytical expression given by Eq.~(55) in Ref.~\cite{GKKM15}, whereas in the region {\bf {\large 1}} $-$ modified expression as defined by Eq.~(\ref{eq:12}).
In the upper row the sum of all contributions is shown. The left (middle) panel corresponds to description of the vector and axial-vector form factors chosen in  Ref.~\cite{GKKM15} which labeled as set 1 (2). The form factors used in the right panel differ by accounting $\rho'$ in the vector form factor and $a_1$ propagator in the axial-vector form factor as explained in text, and we label them as the set 3. In the second (lower) row the separate contributions (the structureless, resonance ones and their interference) are shown for the set 1 (3).}

\end{figure}

Having the model for the resonance contribution we can predict the
branching ratio of the $\tau\rightarrow (a_1\rightarrow \pi^-
+\gamma) +\nu_\tau$ decay estimated experimentally as \cite {ALEPH}
$(4.0\pm 2.0)\cdot 10^{-4}$  .

To obtain the respective number we have to integrate differential
decay width (\ref{eq:12}) over the region the $a_1-$ meson contribution,
namely from $(M_{a_1}-\Gamma_{a_1}/2)^2$ up to
$(M_{a_1}+\Gamma_{a_1}/2)^2$ and divide the result by  the total
$\tau$ lepton width $\Gamma = 2.27\cdot 10^{-12} ~ GeV.$ We
perform two estimations, {\it i)}~integrating all the
contributions in the r.h.s. of Eq.~(\ref{eq:12}) and {\it ii)} leaving the
axial-vector contribution only. For the set 3 we derived the
values $3.8\cdot 10^{-4}$ and $2.8 \cdot 10^{-4},$ respectively
for the branching ratio. For the sets 1 and 2 ($m_{a_1}=1.23~GeV$) we
received the values about three and ten times smaller,
respectively.  When calculating we took
$ \Gamma_{a_1}=0.5~GeV.$

For unpolarized $\tau$ lepton, we define the $t-$ distribution of the photon Stokes parameters $\xi_i(t),\, i=1,\,2,\,3$ as the ratios
of the corresponding partial widths $d\Gamma^0_i(t)/dt$ to $d\Gamma^0(t)/dt.$ The modified expression of the quantity $d\Gamma_{1x}(t,X)/dt$ that defines the parameter $\xi_1(t,X)$ due to the contribution of the region {\bf {\large 1}} reads

\begin{equation}\label{eq:13}
\frac{d\Gamma_{1x}}{d\,t}=P\big[Im(a(t)^*v(t))A_{1x}(t,X)+Im(v(t))C_{1x}(t,X)\big]\,.
\end{equation}
It does not contain the pure IB contribution and it is valid for the choice  of the 4-vectors $e_1^\mu$ and $e_2^\mu$  as defined above.
The simple calculation results
\[A_{1x}(t,X)=-\frac{2 \Omega_x^2 \left[\left(t-m^2\right) \left(M^2-3 t\right)+4 M t X\right]}{3 M^3 t^2}\,,\]
\[C_{1x}(t,X)=\frac{4 \Omega_x \left[\left(m^2-t\right) \left(M^2+2 t\right)+2 M t X\right]}{t \left(m^2-t\right)}-8 M L_x
   \left(m^2-t\right).\]

The modified quantity $d\Gamma_{2x}(t,X)/dt,$ which defines the circular polarization of the photon (parameter $\xi_2(t,X)$) reads
\begin{equation}\label{eq:14}
\frac{d\Gamma_{2x}}{d\,t}=P\big[I_{2x}(t,X)+Re(a(t)^*v(t))A_{2x}(t,X)+[|a(t)|^2+|v(t)|^2]D_{2x}(t,X)
\end{equation}
$$+Re(a(t))B_{2x}(t,X)+Re(v(t))C_{2x}(t,X)\big]\,,$$
where
\[I_{2x}(t,X)=4 M \left[L_x \left(m^2+2 M^2+t\right)+\frac{\Omega_x \left[M X \left(m^2+t\right)+t \left(t-m^2\right)\right]}{t X
   \left(t-m^2\right)}\right]\,,\]
\[A_{2x}(t,X)=-2A_{0x}(t,X)\,, \ B_{2x}(t,X)=-C_{0x}(t,X)\,, \]
\[C_{2x}(t,X)=-B_{0x}(t,X)\,, \ 2D_{2x}(t,X)=-D_{0x}(t,X)\,.\]
The contribution with the factor $[|a(t)|^2+|v(t)|^2]$ vanishes if $X=\omega_{min}$ as it follows from the last relation.

For the part of the differential width connected with the parameter $\xi_3(t,X)$ we have
\begin{equation}\label{eq:15}
\frac{d\Gamma_{3x}}{d\,t}=P\big[I_{3x}(t,X)+\big(|a(t)|^2-|v(t)|^2\big)A_{3x}(t,X)+Re(a(t))B_{3x}(t,X)\big]\,,
\end{equation}
\[I_{3x}(t,X)=\frac{4 M \Omega_x\left(M^2-m^2\right)\left(-m^2+2 M X+t\right)}{X \left(m^2-t\right)^2}-\frac{8 M L_x \left(M^2-m^2\right) \left(M^2+t\right)}{m^2-t}\,,\]
\[2A_{3x}(t,X)=-A_{1x}(t,X)\,, \ B_{3x}(t,X)=-C_{1x}(t,X)\,.\]

The reduced Stokes parameters as a function of the variable $t$ with restriction on the photon energy are shown in Fig.~3.

\begin{figure}

\includegraphics[width=0.32\textwidth]{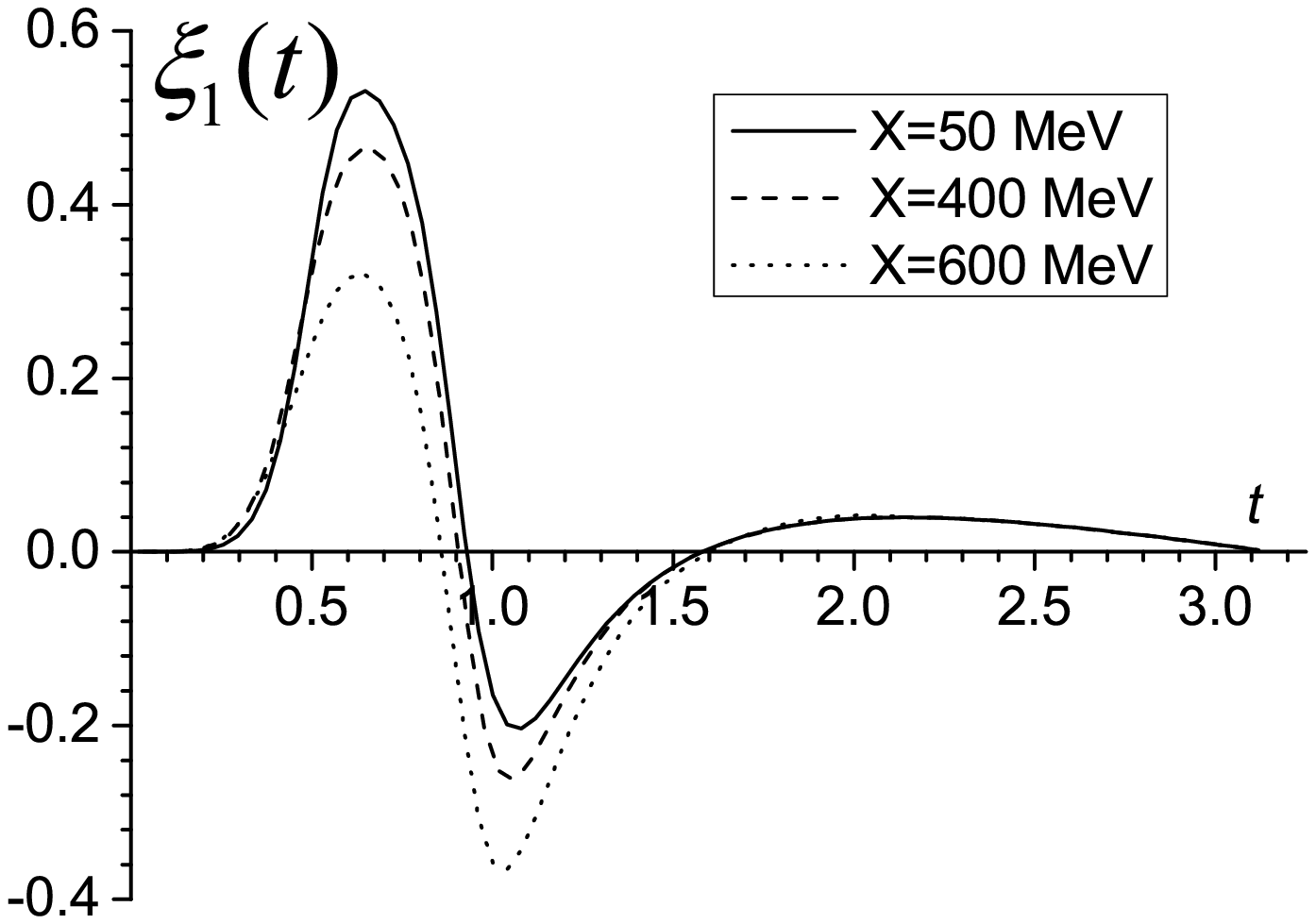}
\includegraphics[width=0.32\textwidth]{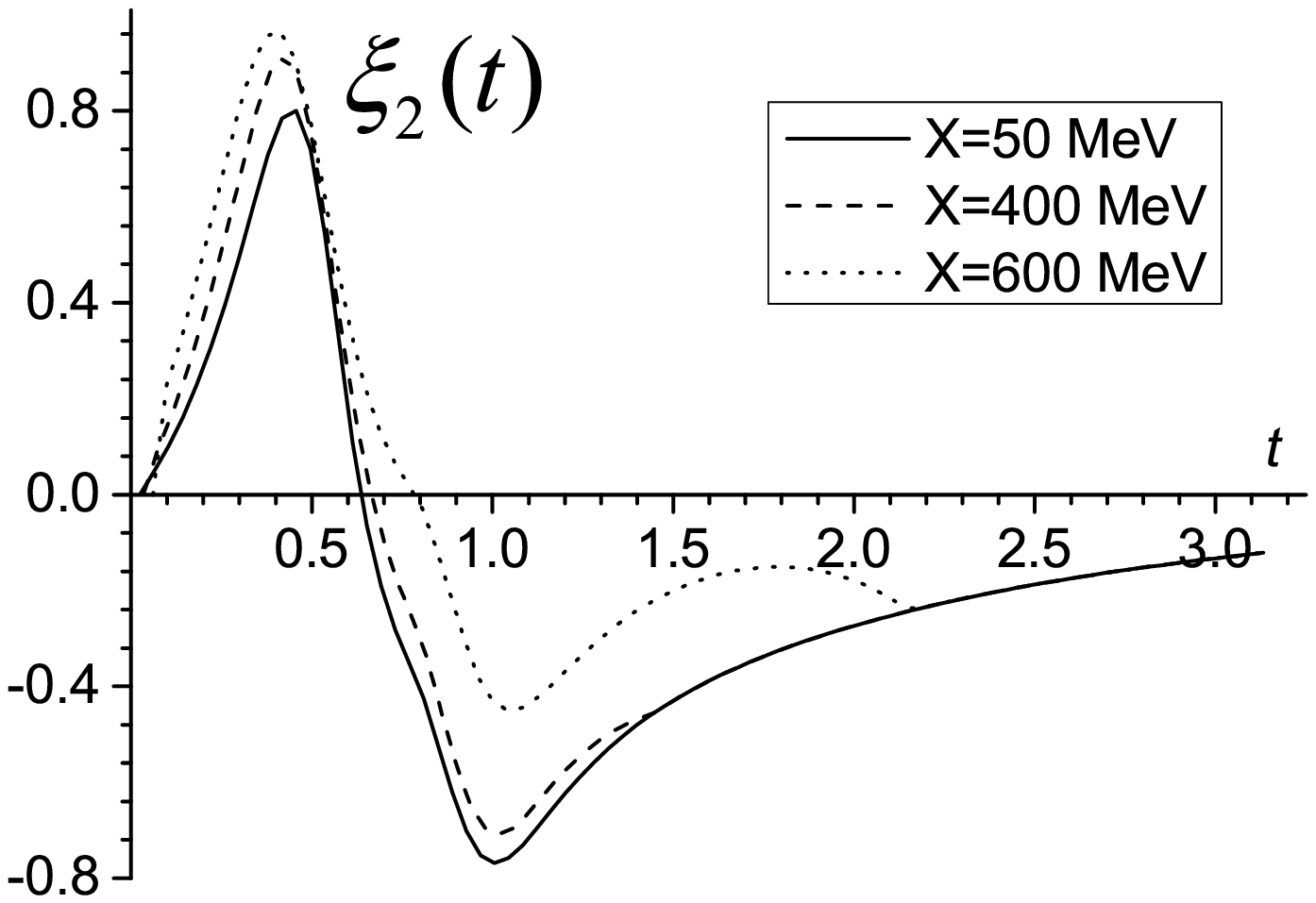}
\includegraphics[width=0.32\textwidth]{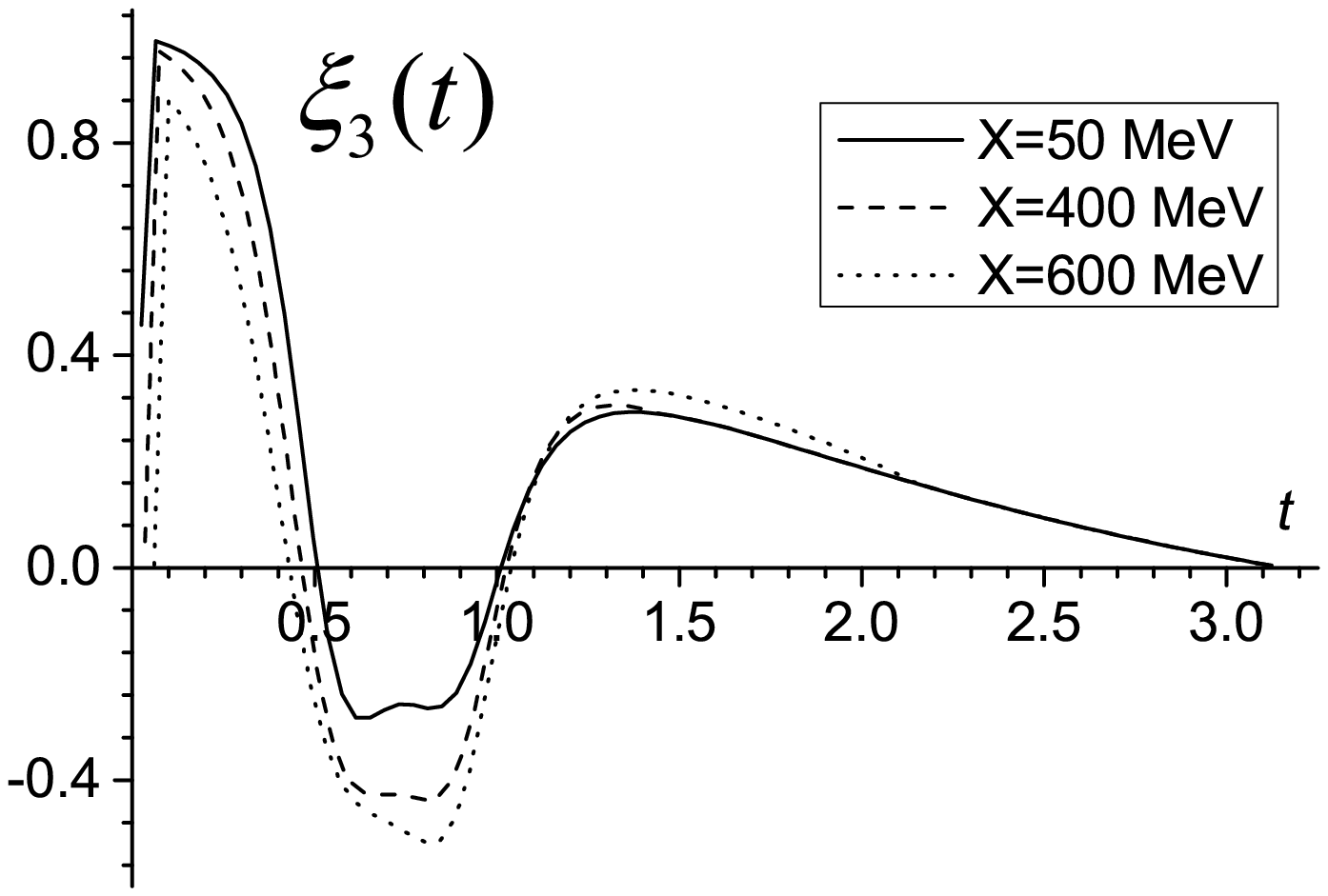}
\vspace{0.5cm}

\includegraphics[width=0.32\textwidth]{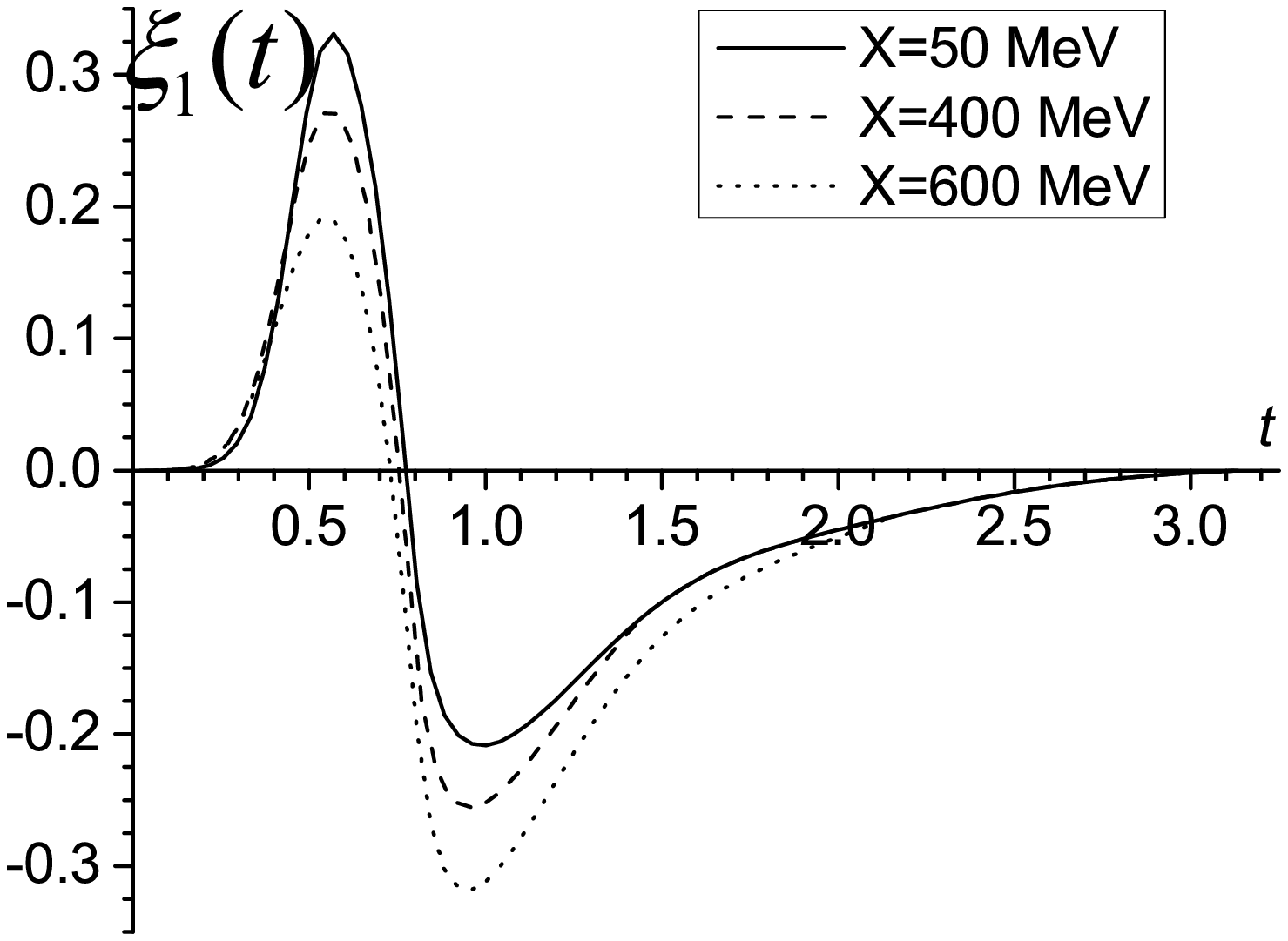}
\includegraphics[width=0.32\textwidth]{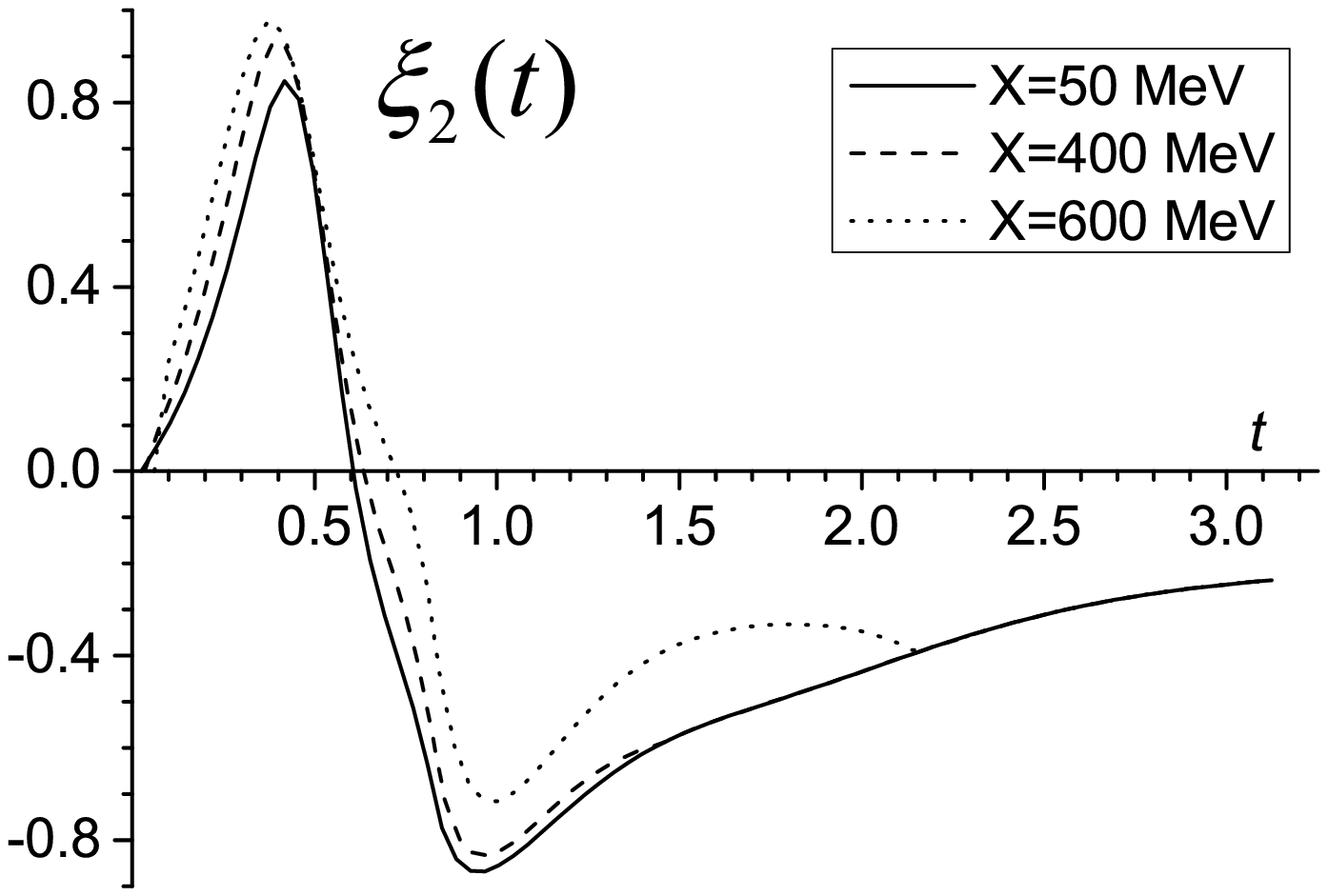}
\includegraphics[width=0.32\textwidth]{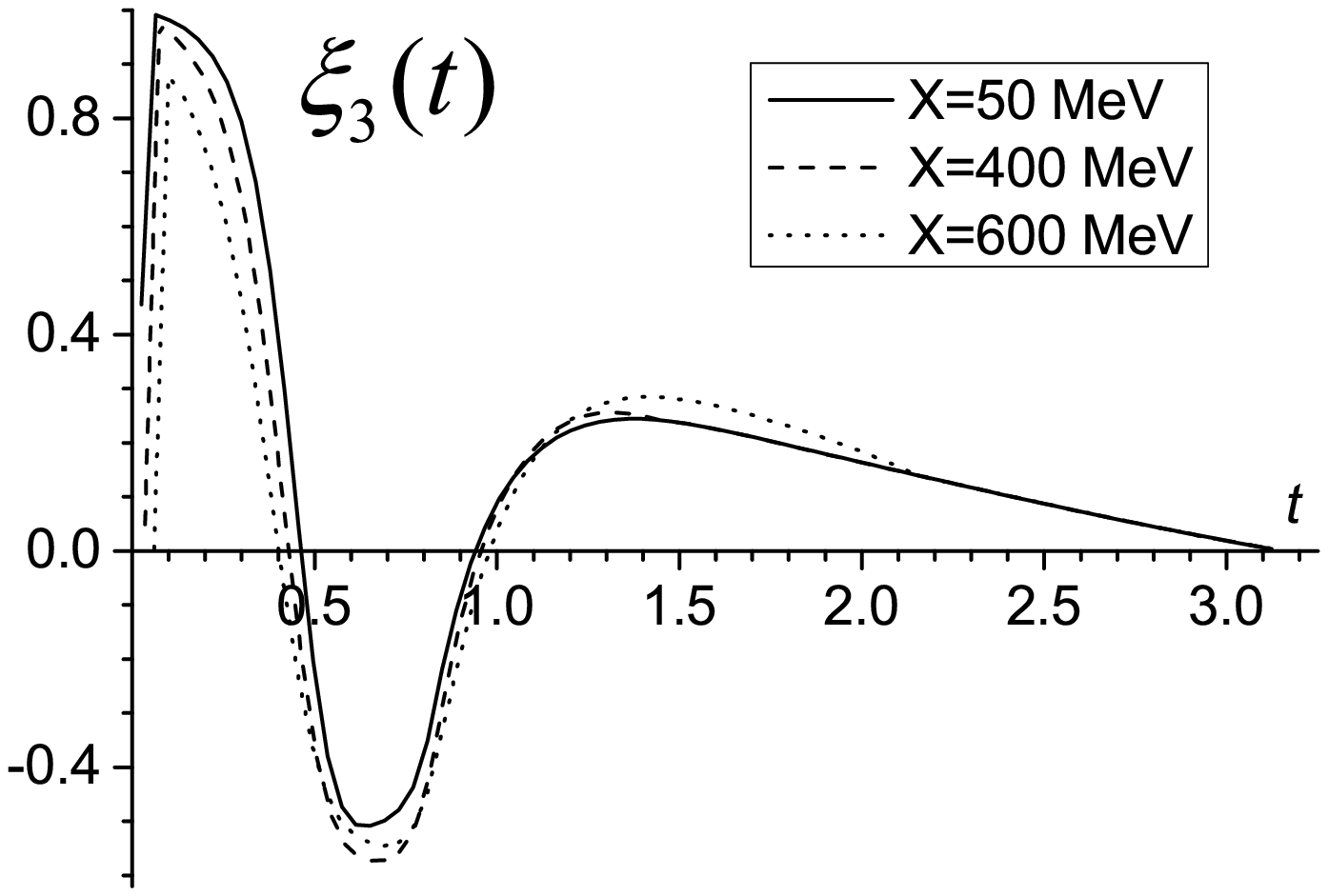}

\caption{The $t-$distribution of the Stokes parameters defined by Eq.~(\ref{eq:6}) for the set 1 used in Ref.~\cite{GKKM15} (upper row) and set 3 (lower row).
The parameters $\xi_1$ and $\xi_3$, which describe the linear polarization of the photon,
 are determined relative to the orthogonal planes (${\bf k,\,q}$) and (${\bf e_2,\,k}$) in the rest frame of the $\tau$ lepton.}
\end{figure}

Let us consider the P-odd effects due to the terms containing $(Sq)$ and $(Sk)$ in the matrix element squared in the case of polarized $\tau$ lepton.
The corresponding differential decay width contains polarization-independent and polarization-dependent parts, that is proportional to $c_2$, and can be written as
\begin{equation}\label{eq:16}
\frac{d\,\overline{\Gamma}_0}{d\,t\,dc_2}=\frac{1}{2}\Big[\frac{d\,\Gamma_0(t)}{d\,t d c_2}+\frac{c_2\,d\,\Gamma_0^{^S}(t)}{d\,t d c_2}\Big]\,.
%\frac{d\,\Gamma_0^{^S}(t,c_2)}{dc_2\,dt}=\frac{1}{2}\frac{d\,\overline{\Gamma}_0^{^S}(t)\,c_2}{dc_2\,dt}
\end{equation}
The elementary angular integration in the upper and down hemispheres gives
$$\frac{d\,\overline{\Gamma}_0^{(up)}}{d\,t}=\frac{1}{2}\Big[\frac{d\,\Gamma_0(t)}{d\,t}+\frac{1}{2}\frac{d\,\Gamma_0^{^S}(t)}{d\,t}\Big]\,, \ \
\frac{d\,\overline{\Gamma}_0^{(dn)}}{d\,t}=\frac{1}{2}\Big[\frac{d\,\Gamma_0(t)}{d\,t}-\frac{1}{2}\frac{d\,\Gamma_0^{^S}(t)}{d\,t}\Big]\,.$$

Now we define the {\it up-down} asymmetry as the ratio
\begin{equation}\label{eq:17}
A^{ud}(t)=\frac{d\,\overline{\Gamma}_0^{(up)}/dt-d\,\overline{\Gamma}_0^{(dn)}/dt}{d\,\overline{\Gamma}_0^{(up)}/dt+d\,\overline{\Gamma}_0^{(dn)}/dt}=
\frac{1}{2}\frac{d\,\Gamma_0^{^S}(t/d t)}{d\,\Gamma_0(t)/d t}\,,
\end{equation}
where $d\,\Gamma_0(t)/d t$ is defined by Eq.~(\ref{eq:12}) in the region {\large{\bf 1}} and by Eq.~(55) of the Ref.~\cite{GKKM15} in the region {\large{\bf 2}}. As concerns the polarization-dependent part of the decay width in the region {\large{\bf 1}} it reads

\begin{equation}\label{eq:18}
\frac{d\Gamma^{^S}_0(t,X)}{d\,t}=P\big[I^{^S}_0(t,X)+\big(|a(t)|^2+|v(t)|^2\big)A^{^S}_0(t,X)+
\end{equation}
$$Re(a(t)^*v(t))B^{^S}_0(t,X)+Re(a(t))C^{^S}_0(t,X)+Re(v(t))D^{^S}_0(t,X)\big]\,,$$
\[I^{^S}_0(t,X)=\frac{2 M \Omega_x \left[2 M X^2 \left(m^4+t \left(2 M^2+t\right)\right)+2 t X \left(t-m^2\right) \left(m^2+3
   M^2+t\right)-M t \left(m^2-t\right)^2\right]}{t X^2 \left(m^2-t\right)^2}\]
   \[-\frac{4 M L_x \left(m^4+2 m^2 M^2+2 M^4+6 M^2
   t+t^2\right)}{m^2-t}\,,\]
\[A^{^S}_0(t,X)=-\frac{\Omega_x \left[\left(m^2-t\right)^2 \left(M^4-6 M^2 t-3 t^2\right)-2 M t X
   \left(\left(m^2-t\right) \left(M^2+3 t\right)+4 M t X\right)\right]}{3 M^4 t^2}\]
\[-\frac{2 L_x \left(m^2-t\right)^3}{M^3}\,,\]
\[B^{^S}_0(t,X)=\frac{4 L_x\left(m^2-t\right)^3}{M^3}+\frac{2 \Omega_x \left(t-m^2\right)\left[\left(m^2-t\right) \left(2
   M^2+t\right)+2 M t X\right]}{M^4 t}\,,\]
\[C^{^S}_0(t,X)=-\frac{4\Omega_x \left[M X \left(m^2-t\right) \left(m^2+M^2+3
   t\right)-t \left(m^2-t\right)^2+2 M^2 t X^2\right]}{M t X \left(m^2-t\right)}\]
\[+\frac{4 L_x \left(m^2-t\right) \left(m^2+4 M^2+t\right)}{M}\,,\]
\[D^{^S}_0(t,X)=\frac{4 \Omega_x \left[M X \left(m^2+t\right)+t \left(t-m^2\right)\right]}{M t X}-\frac{4 L_x\left(m^2-t\right)
   \left(m^2+2 M^2+t\right)}{M}\,.\]
In the region {\large{\bf 2}} the corresponding expressions for $I^{^S}_0(t),\,\,A^{^S}_0(t),\,\,B^{^S}_0(t),\,\,C^{^S}_0(t)$ and $D^{^S}_0(t)$
are given by the relations (59) of Ref.~\cite{GKKM15}.

The similar procedure can be applied to extract the polarization-dependent parts of the Stokes parameters (the correlation parameters) $\xi_i^{(ud)}$
keeping in mind that the quantities $d\Gamma_i^{^S}(t,c_2)/dt dc_2$ are also proportional to $c_2.$ Provided that the normalization of the correlation parameters
is carried out by the polarization-independent part of the decay width, we can write by analogy with Eq.~(\ref{eq:17})
\begin{equation}\label{eq:19}
\xi_i^{ud}(t)=\frac{1}{2}\frac{d\,\Gamma_i^{^S}(t/d t)}{d\,\Gamma_0(t)/d t}\,.
\end{equation}

In the region {\large{\bf 1}}, we have
for the quantities connected with the correlation parameters the result
\begin{equation}\label{eq:20}
\frac{d\Gamma^{^S}_1(t,X)}{dt}=P\big[Im(a(t)^*v(t))B^{^S}_1(t,X)+Im(a(t))C^{^S}_1(t,X)+Im(v(t))D^{^S}_1(t,X)\big]\,,
\end{equation}
\[B^{^S}_1(t,X)=A_1(t,X)\,, \ C^{^S}_1(t,X)=\frac{8 L_x \left(m^2-t\right) \left(M^2+t\right)}{M}-\frac{4 \Omega_x \left(-m^2+2 M X+t\right)}{M X}\,,\]
\[D^{^S}_1(t,X)=\frac{4 \Omega_x \left[2M^2 X^2 t+M X(m^2-t)(M^2+4t)-t(m^2-t)^2\right]}{M X t(m^2-t)}
-\frac{8 L_x \left(m^2-t\right)(2M^2+t)}{M}\,,\]

\begin{equation}\label{eq:21}
\frac{d\Gamma^{^S}_2(t,X)}{dt}=P\big[I^{^S}_2(t,X)+(|a(t)|^2+|v(t)|^2\big)A^{^S}_2(t,X)
+Re(a(t)^*v(t))B^{^S}_2(t,X)+
\end{equation}
$$Re(a(t))C^{^S}_2(t,X)+Re(v(t))D^{^S}_2(t,X)\big]\,,$$
\[I^{^S}_2(t,X)=\frac{M^2}{m^2-t}D^{^S}_0(t,X)\,, \ B^{^S}_2(t,X)=-2A^{^S}_0(t,X)\,,\]
\[  C^{^S}_2(t,X)=-D^{^S}_0(t,X)\,, \ D^{^S}_2(t,X)=-C^{^S}_0(t,X)\,,\]
\[A^{^S}_2(t,X)=\frac{\left(m^2-t\right) \Omega_x \left[\left(m^2-t\right) \left(2 M^2+t\right)+2 M t X\right]}{M^4 t}-\frac{2 L_x
   \left(m^2-t\right)^3}{M^3}\,,\]

\begin{equation}\label{eq:22}
\frac{d\Gamma^{^S}_3(t,X)}{dt}=P\big[I^{^S}_3(t,X)+(|a(t)|^2-|v(t)|^2\big)A^{^S}_3(t,X)
+Re(a(t))C^{^S}_3(t,X)+Re(v(t))D^{^S}_3(t,X)\big]\,,
\end{equation}
\[I^{^S}_3(t,X)=\frac{8 M L_x \left[m^2 \left(M^2+t\right)+M^2 \left(M^2+3 t\right)\right]}{t-m^2}\]
\[+\frac{2 M \Omega_x \left[-2 X
   \left(m^2-t\right) \left(m^2+3 M^2+t\right)+4 M X^2 \left(m^2+M^2\right)-M \left(m^2-t\right)^2\right]}{X^2
   \left(m^2-t\right)^2}\,,\]
\[A^{^S}_3(t,X)=-\frac{1}{2}A_1(t,X)\,, \ C^{^S}_3(t,X)=-D^{^S}_1(t,X)\,, \ D^{^S}_3(t,X)=-C^{^S}_1(t,X)\,.\]

The corresponding numerical results are illustrated in Fig.~4, where we show the $t$-distributions of the {\it up-down} asymmetry and the correlation
parameters for different photon energy cuts.

\begin{figure}

\includegraphics[width=0.23\textwidth]{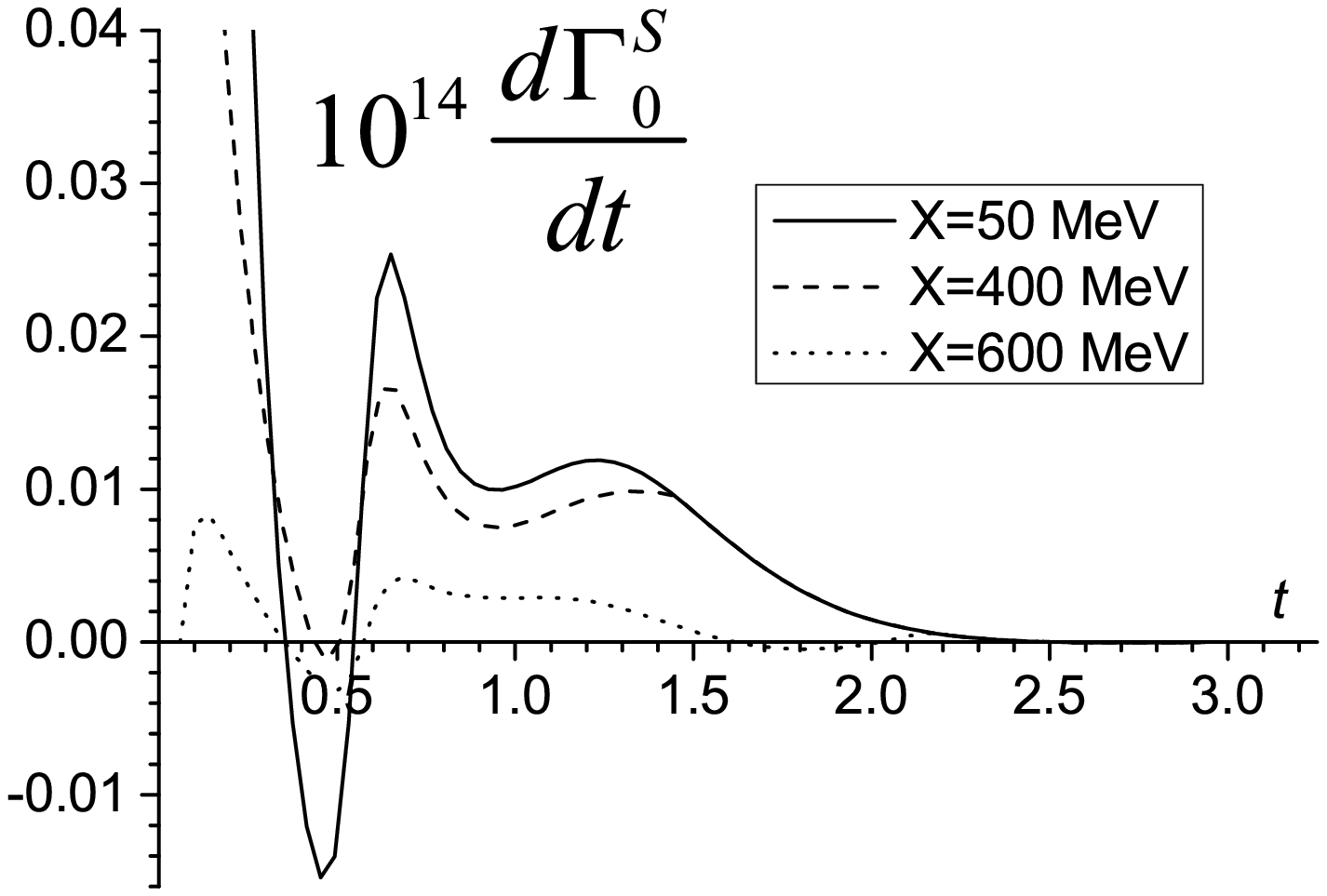}
\includegraphics[width=0.23\textwidth]{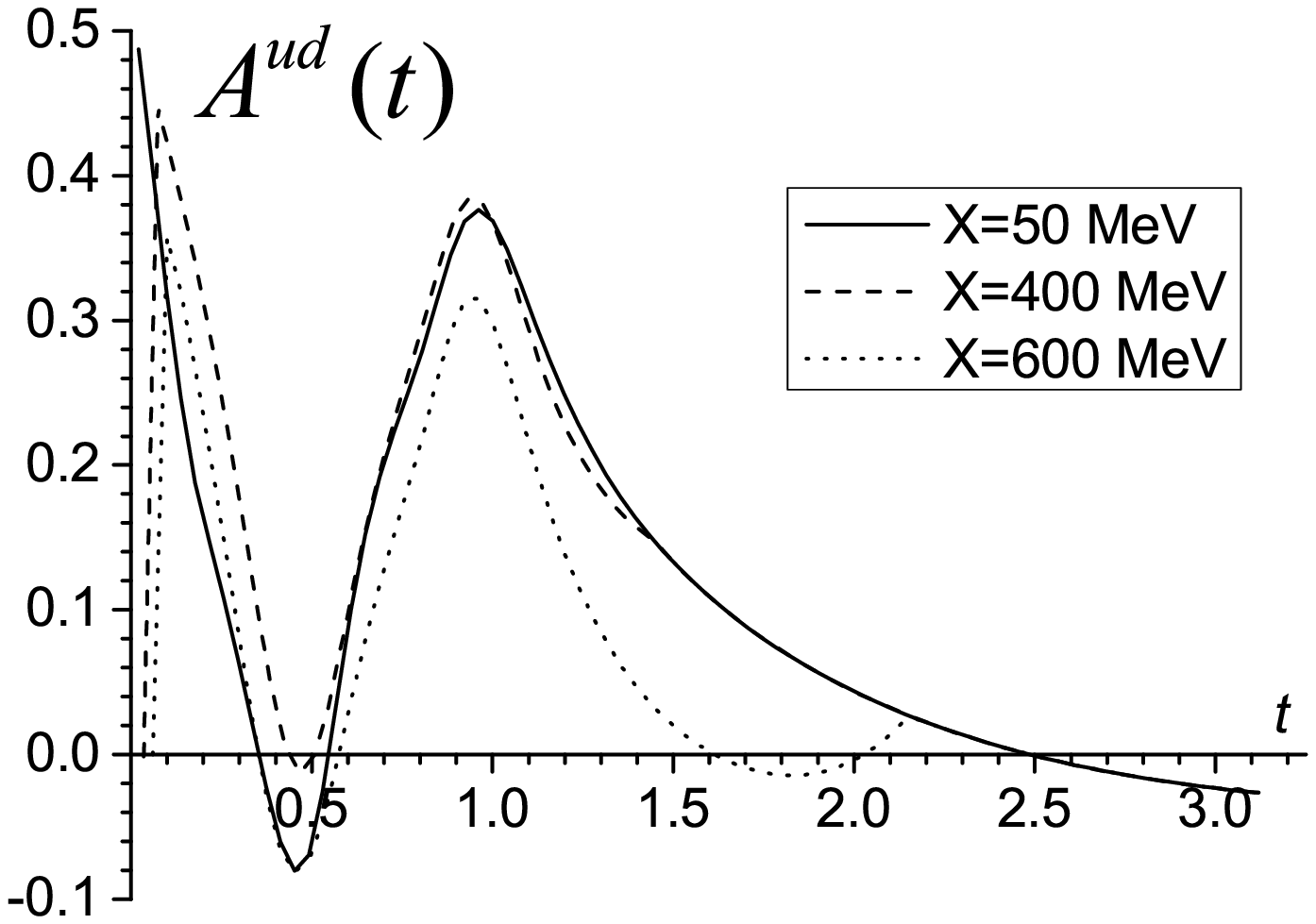}
\includegraphics[width=0.23\textwidth]{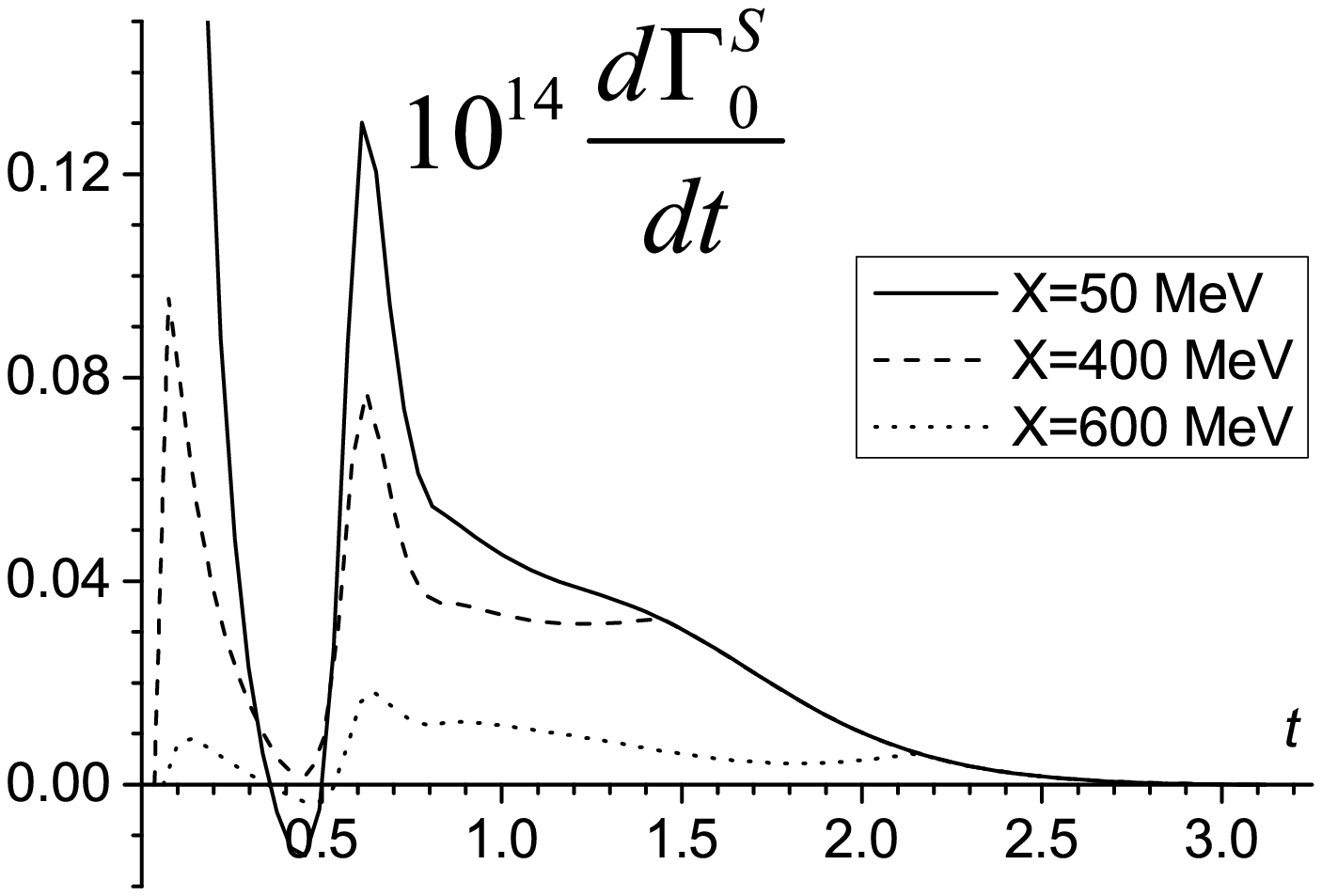}
\includegraphics[width=0.23\textwidth]{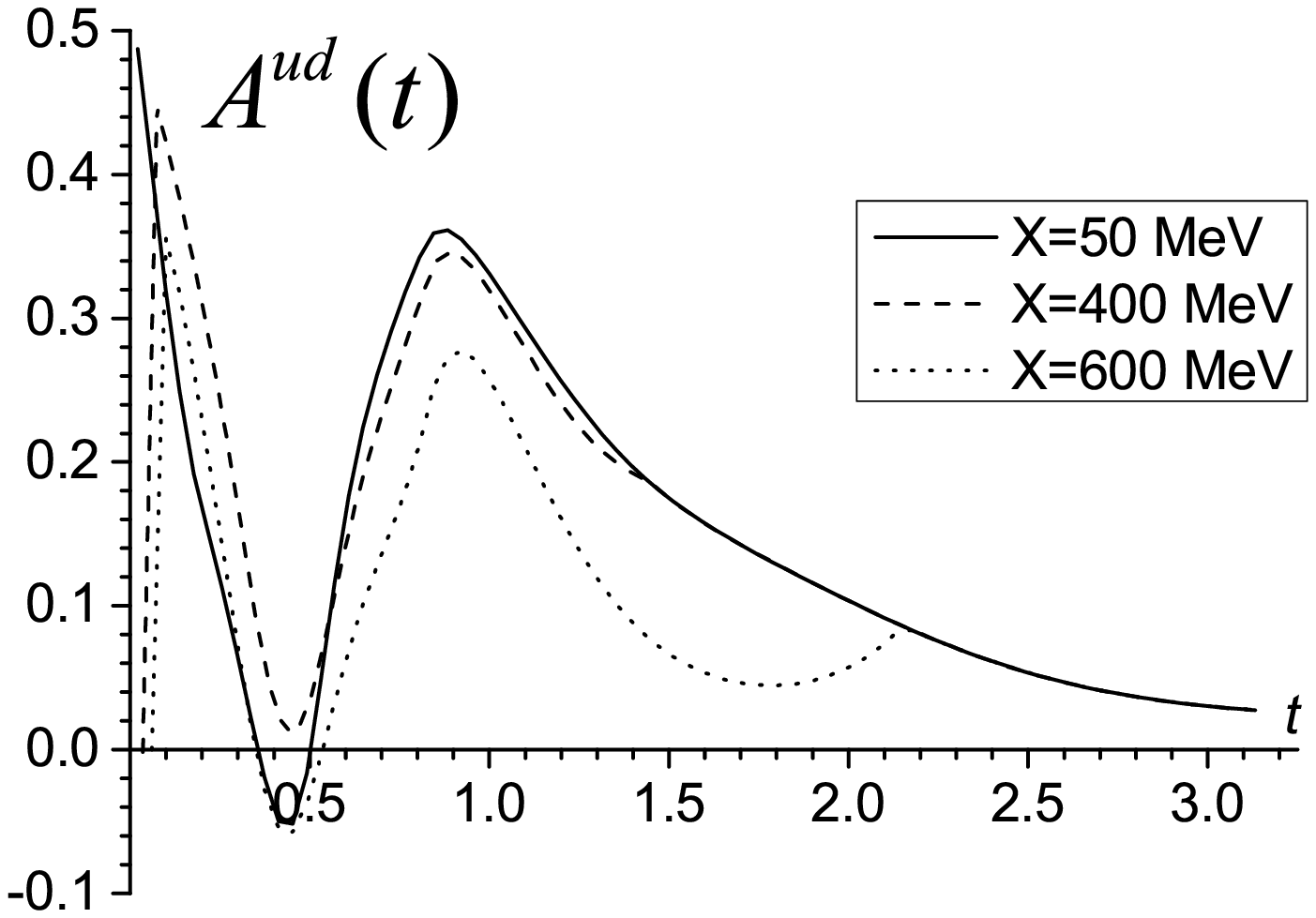}
\vspace{0.4cm}

\includegraphics[width=0.30\textwidth]{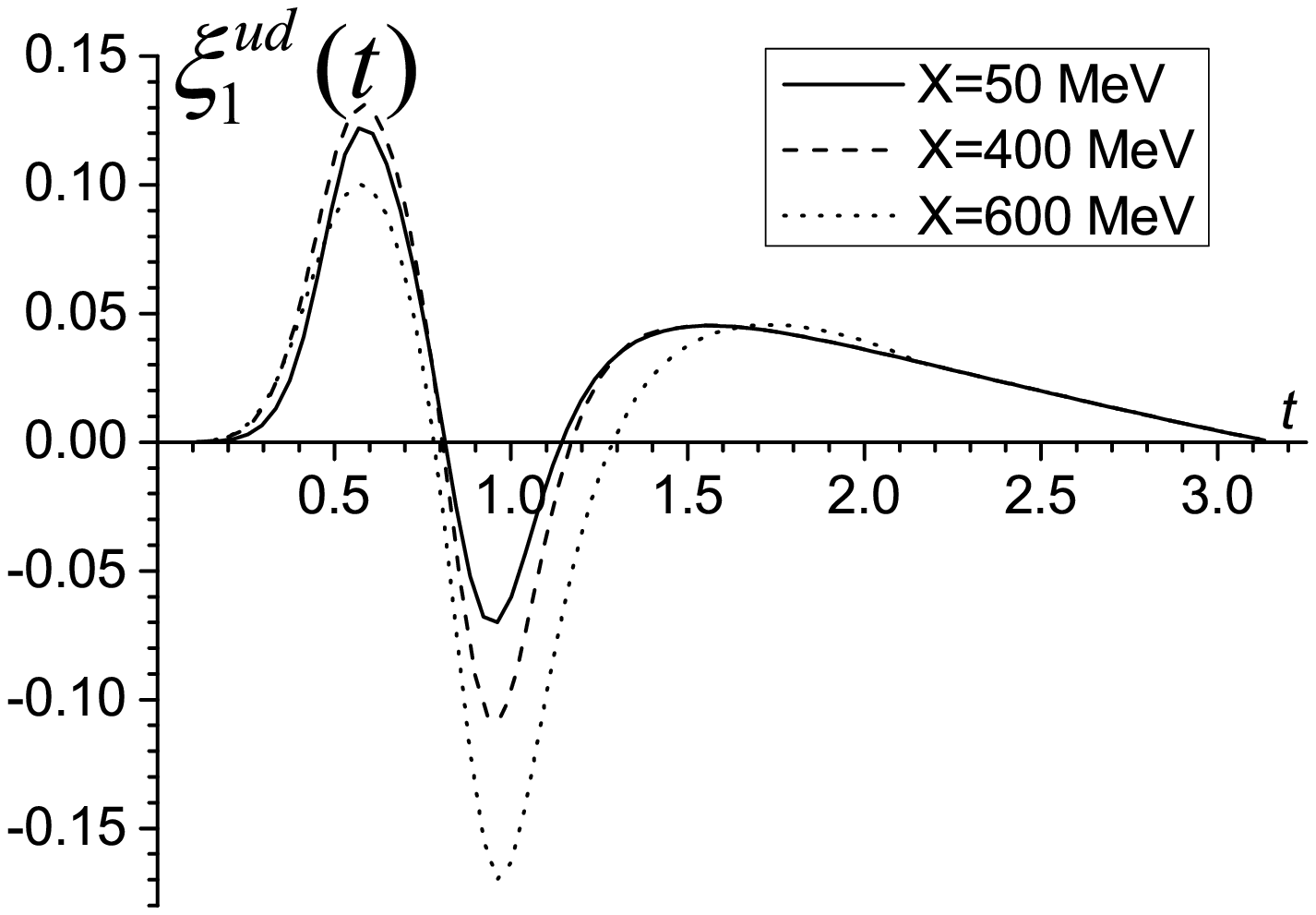}
\includegraphics[width=0.30\textwidth]{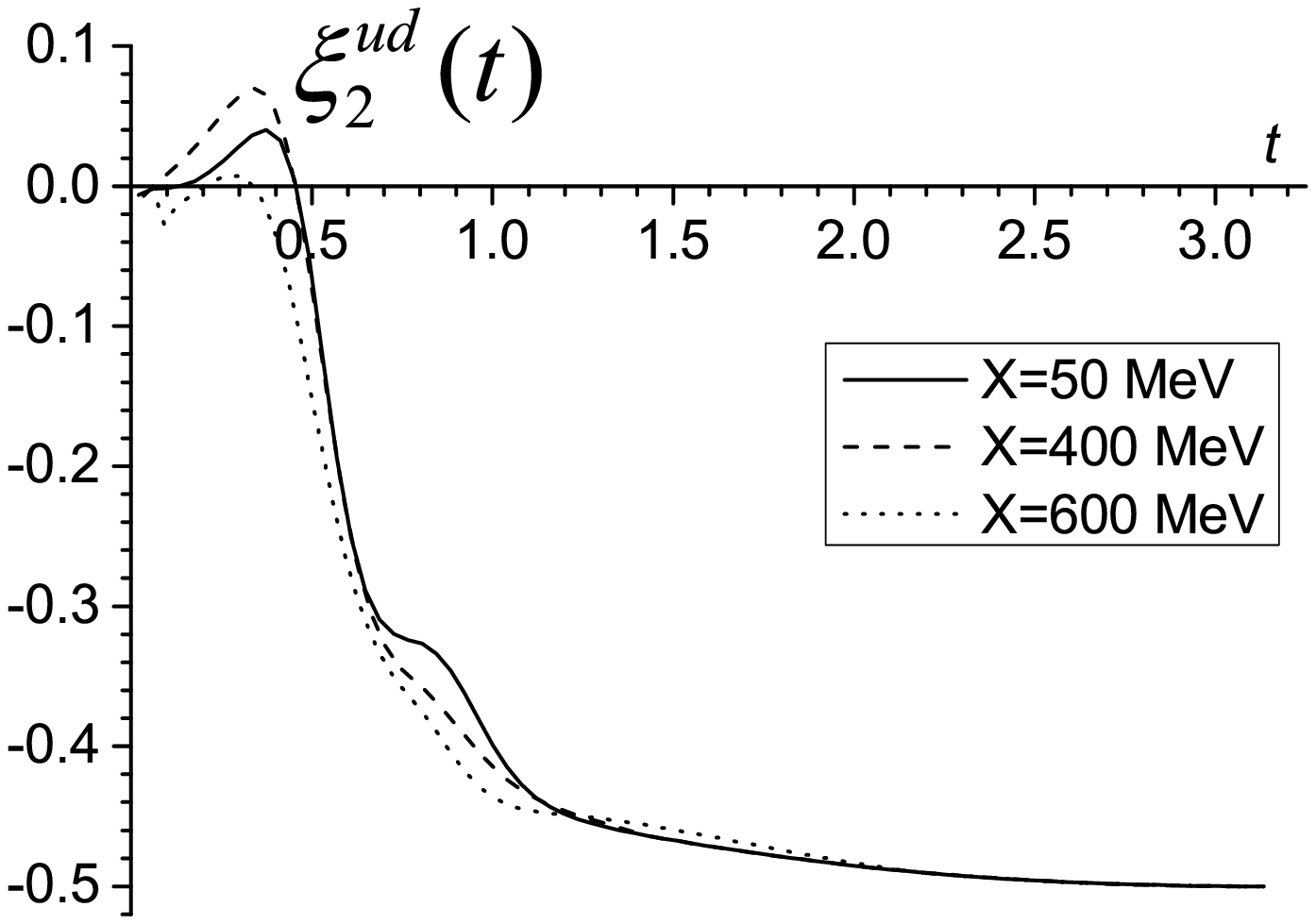}
\includegraphics[width=0.30\textwidth]{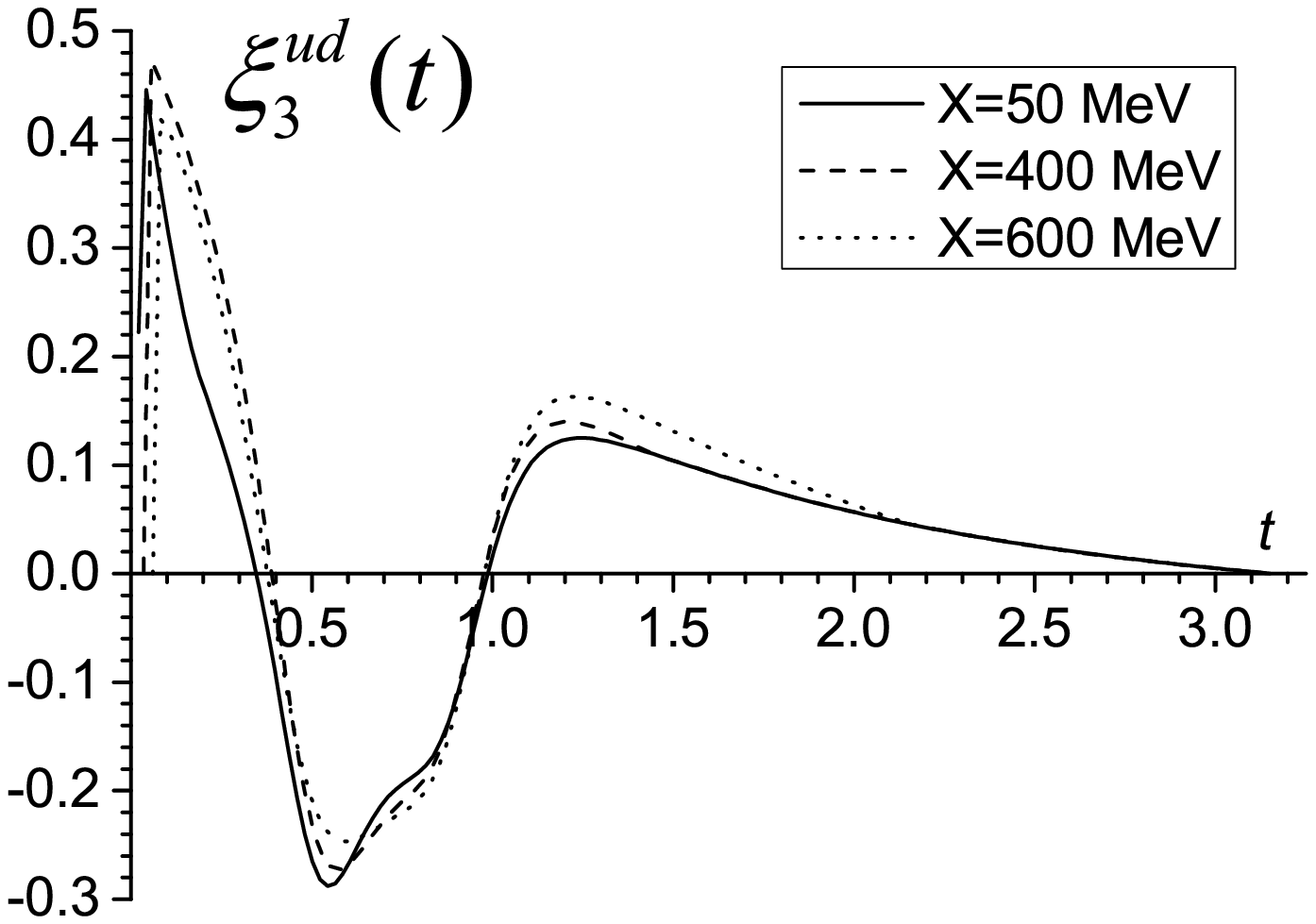}
\vspace{0.4cm}

\includegraphics[width=0.30\textwidth]{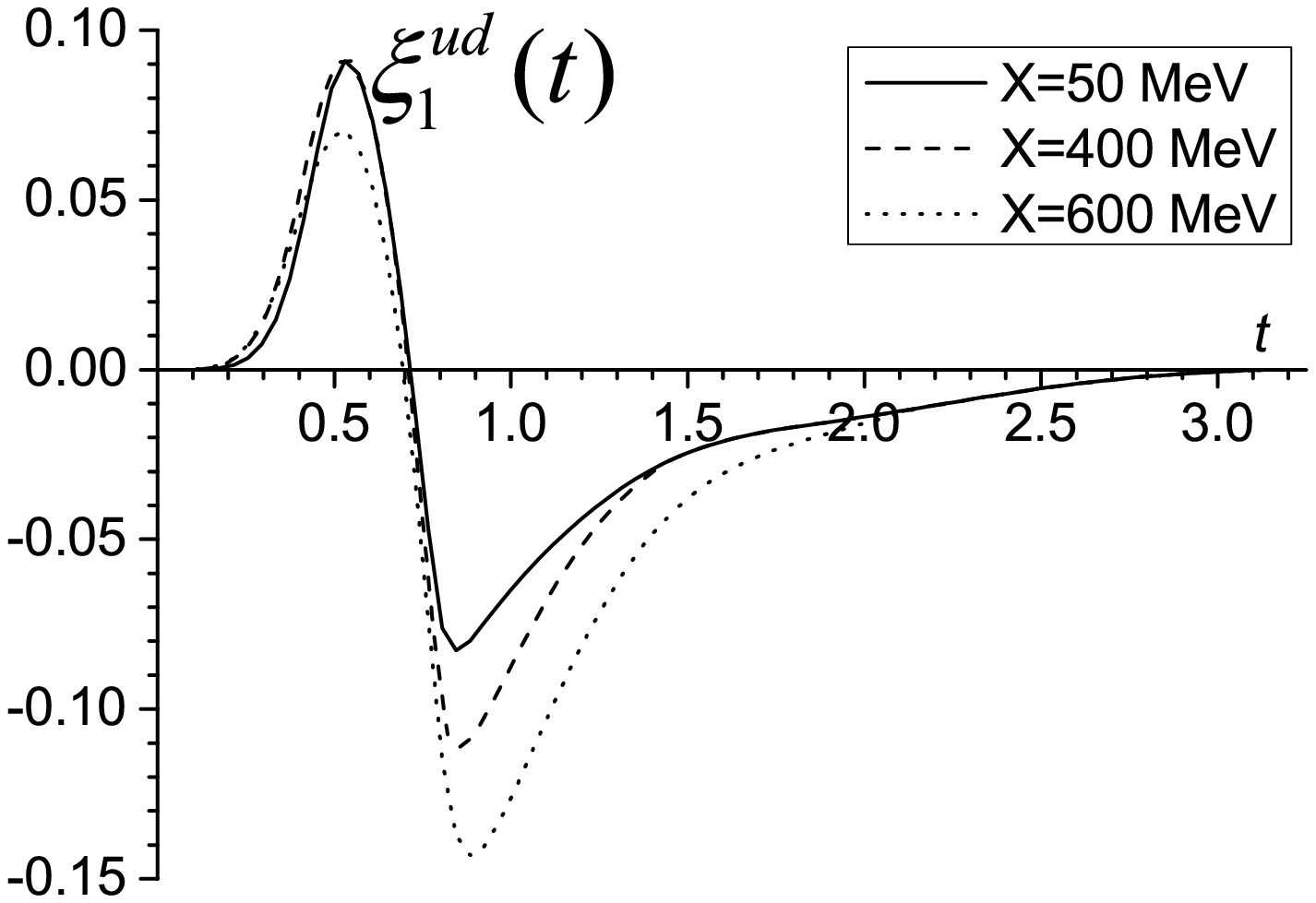}
\includegraphics[width=0.30\textwidth]{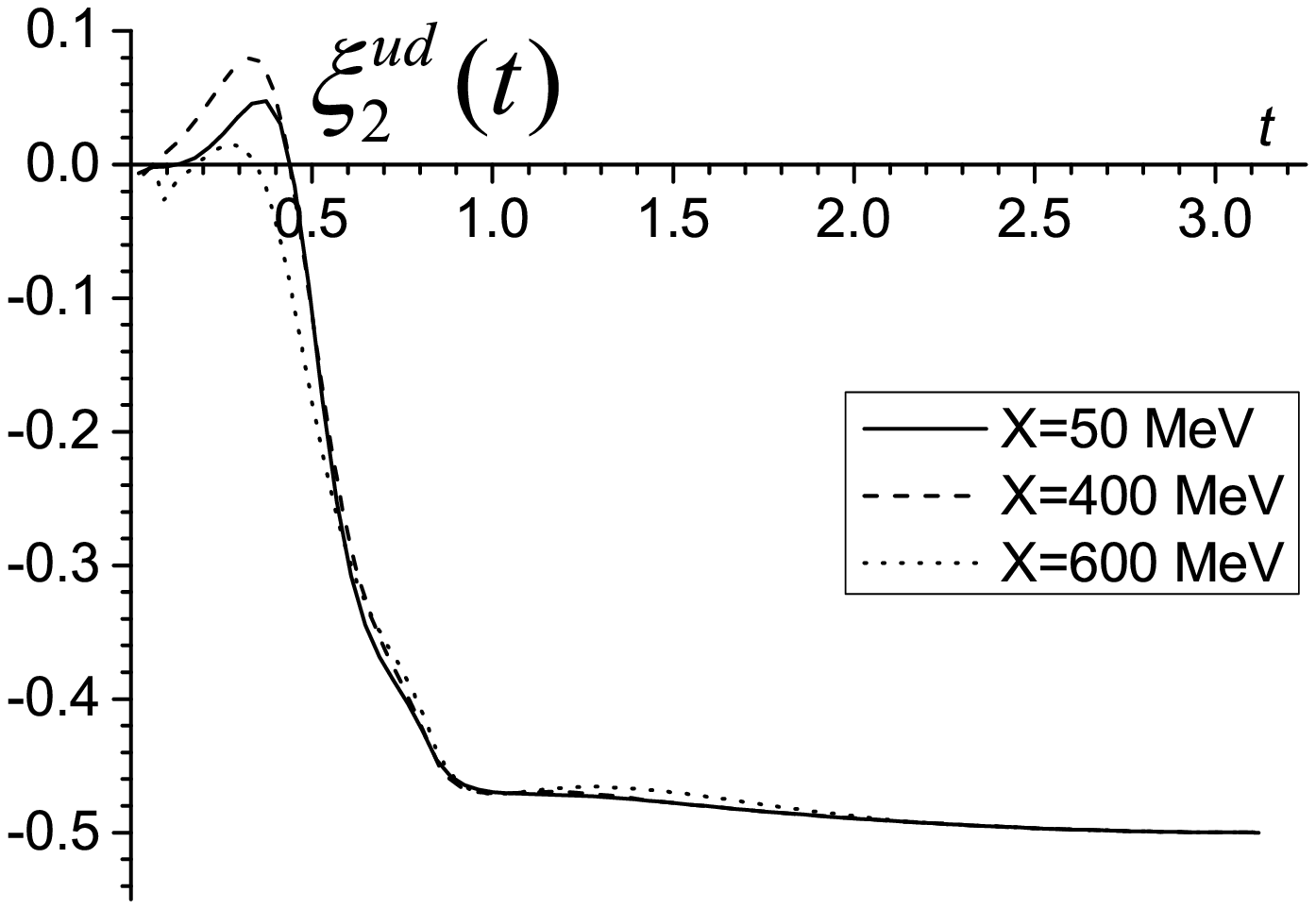}
\includegraphics[width=0.30\textwidth]{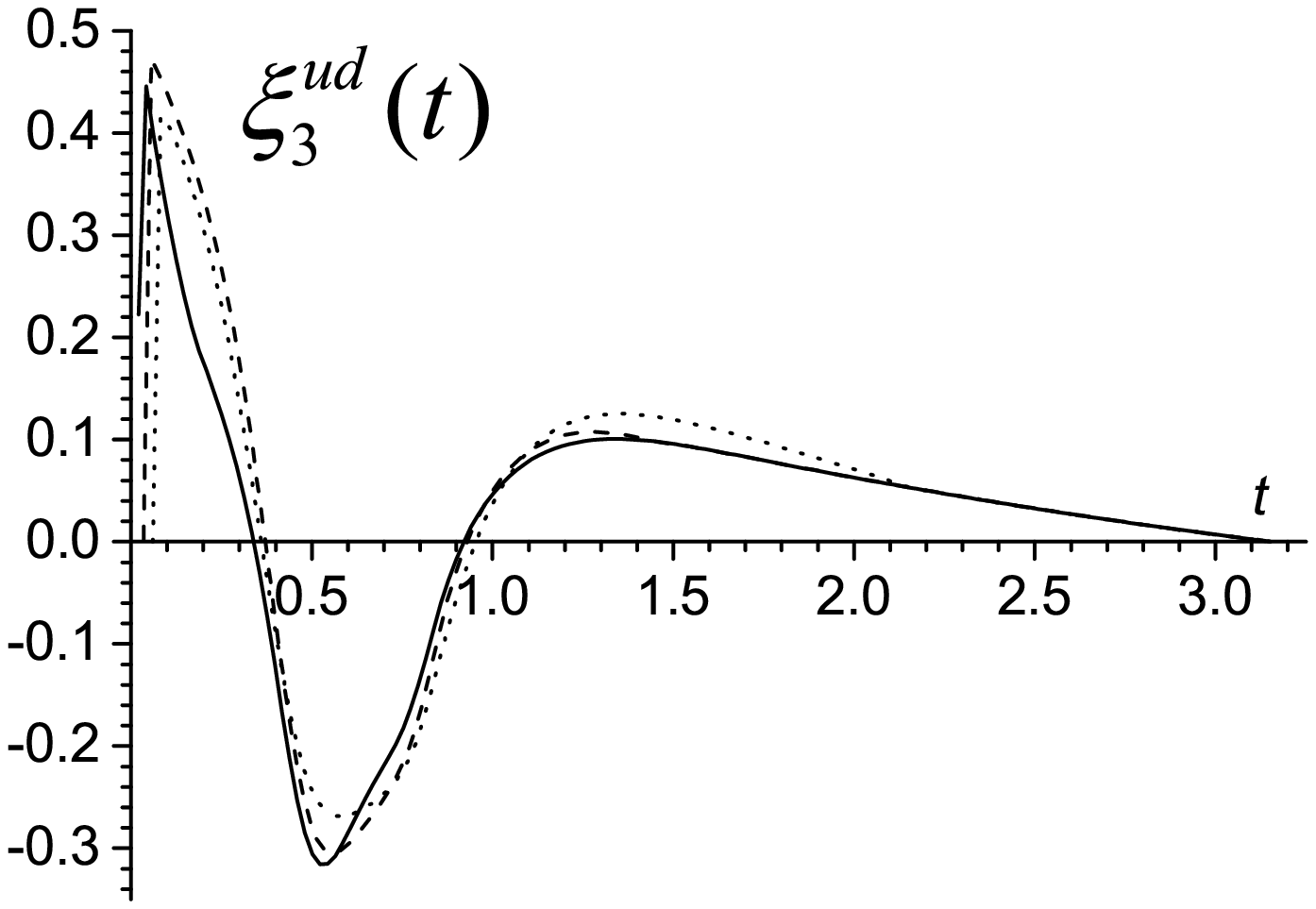}

\caption{The $t-$distribution of the {\it up-down} decay width, the polarization asymmetry and corresponding correlation parameters defined by Eq.~(\ref{eq:17}) for the set 1 used in Ref.~\cite{GKKM15} (the first two panels in the upper row and the second row) and
for the set 3 (the last two panels in the upper row and the lower row). All these quantities arise due to the P-odd terms (Sk) and (Sq) in the matrix element squared.}
\end{figure}

Up to now we considered the P-odd polarization effects caused by the terms $(Sq)$ and $(Sk)$ in the matrix element squared.
To describe the P-even effects, we have to analyse the contribution caused by the term $(Spqk)$ in the matrix element squared. This contribution is
proportional to $\sin{\phi}$ and have the opposite sings in the right ($0<\phi<\pi$) and left ($\pi<\phi<2\pi$) hemispheres. Therefore, the P-even effects
can manifest themselves as the difference of the events in these hemispheres. We denote the corresponding {\it right-left} asymmetry and the correlation
parameters as $A^{^{RL}}(t)$ and $\xi_i{^{RL}}(t),$  respectively. By analogy with the corresponding {\it up-down} quantities, they can be
written as
\begin{equation}\label{eq:23}
A^{^{RL}}(t)=\frac{1}{2}\frac{d\,\Gamma_0^{^{RL}}(t/d t)}{d\,\Gamma_0(t)/d t}\,, \  \xi_i{^{RL}}(t)
=\frac{1}{2}\frac{d\,\Gamma_i^{^{RL}}(t/d t)}{d\,\Gamma_0(t)/d t}\,,
\end{equation}
In the region {\large{\bf 2}} when $X=t_-,$ all these quantities are given by Eqs.~(12-15) in Ref.~\cite{GKM16}

To find these quantities for the restricted photon phase space, we have to calculate the following integrals (see also Eq.~(\ref{eq:11}))
\begin{equation}\label{eq:24}
J_{0x}(t,X)=\int\limits_X^{\omega_+}|{\bf{q}}|\omega s_{12}\,d\omega=\frac{\sqrt{t}}{8}\Big\{(\omega_+-\omega_-)^2\Big(\frac{\pi}{2}
+\arcsin{\Big[\frac{\omega_+ + \omega_- -2X}{\omega_+-\omega_-}\Big]}\Big)
\end{equation}
\[+2(\omega_+ + \omega_--2X)R \Big\}\,, \ R=\sqrt{(X-\omega_-)(\omega_+-X)}\,,\]

\begin{equation}\label{eq:25}
J_{1x}(t,X)=\int\limits_X^{\omega_+}|{\bf{q}}|s_{12}\,d\omega=\frac{\sqrt{t}}{2}\Big\{(\sqrt{\omega_+}-\sqrt{\omega_-})^2\frac{\pi}{2}+
(\omega_++\omega_-)\arcsin{\Big[\frac{\omega_+ + \omega_- -2X}{\omega_+-\omega_-}\Big]}
\end{equation}
\[+2\sqrt{\omega_+\omega_-}\arcsin{\Big[\frac{X(\omega_+ + \omega_-) -2\omega_+\omega_-}{X(\omega_+-\omega_-)}\Big]}-2R\Big\}\,,\]

\begin{equation}\label{eq:26}
J_{2x}(t,X)=\int\limits_X^{\omega_+}\frac{|{\bf{q}}|s_{12}}{\omega}\,d\omega=\frac{\sqrt{t}}{2\sqrt{\omega_+\omega_-}}\Big\{
(\sqrt{\omega_+}-\sqrt{\omega_-})^2\frac{\pi}{2}-(\omega_+ + \omega_-)\arcsin{\Big[\frac{X(\omega_+ + \omega_-) -2\omega_+\omega_-}{X(\omega_+-\omega_-)}\Big]}
\end{equation}
\[-2\sqrt{\omega_+\omega_-}\arcsin{\Big[\frac{\omega_+ + \omega_- -2X}{\omega_+-\omega_-}\Big]}+\frac{2R\sqrt{\omega_+\omega_-}}{X}\Big\}\,.\]

In terms of these integrals we have
\begin{equation}\label{eq:27}
\frac{d\,\Gamma^{^{RL}}_{0x}}{d\,t}=P\big[Im(a(t))C^{^{RL}}_{0x}(t,X)+
Im(v(t))D^{^{RL}}_{0x}(t,X)\big]\,,
\end{equation}
\[C^{^{RL}}_{0x}(t,X)=\frac{-8\,[J_{1x}(m^2+M^2)(m^2-t)+2MJ_{0x}(m^2+t)]}{M(m^2-t)}\,,\]
\[D^{^{RL}}_{0x}(t,X)=8\,\Big[2J_{0x}-\frac{J_{1x}(M^2-m^2)}{M}\Big]\]
for the quantity that defines the right-left polarization asymmetry. The rest quantities  $d\Gamma_{ix}^{^{RL}}/{dt}$ are defined as
\begin{equation}\label{eq:28}
\frac{d\,\Gamma^{^{RL}}_{1x}}{d\,t}=P\big[I^{^{RL}}_{1x}(t,X)
+\big(|a(t)|^2-|v(t)|^2\big)A^{^{RL}}_{1x}(t,X)+Re(a(t))C^{^{RL}}_{1x}(t,X)+
Re(v(t))D^{^{RL}}_{1x}(t,X)\big]\,,
\end{equation}
\[I^{^{RL}}_{1x}(t,X)=8M^2\,\Big[J_{2x}-\frac{2M\,J_{1x}}{t-m^2}\Big]\,, \ \ A^{^{RL}}_{1x}(t,X) =-\frac{8\,J_{0x}(t-m^2)}{M^2}\,,\]
\[C^{^{RL}}_{1x}(t,X)=-8\,\Big[2\,J_{0x}+\frac{J_{1x}(M^2-t)}{M}\Big]\,, \
\ D^{^{RL}}_{1x}(t,X)=-8\,\Big[\frac{2\,J_{0x}(m^2+t)}{m^2-t}+\frac{J_{1x}(M^2+t)}{M}\Big]\,,\]
\begin{equation}\label{eq:29}
\frac{d\,\Gamma^{^{RL}}_{2x}}{d\,t}=P\big[Im(a(t))C^{^{RL}}_{2x}(t,X)+
Im(v(t))D^{^{RL}}_{2x}(t,X)\big]\,,
\end{equation}
\[C^{^{RL}}_{2x}(t,X)=-D^{^{RL}}{0x}(t,X)\,, \ \ D^{^{RL}}_{2x}(t,X)=-C^{^{RL}}_{0x}(t,X)\,,\]
\begin{equation}\label{eq:30}
\frac{d\,\Gamma^{^{RL}}_{3x}}{d\,t}=P\big[Im(a^*(t)v(t))B^{^{RL}}_{3x}(t,X)+
Im(a(t))C^{^{RL}}_{3x}(t,X)+Im(v(t))D^{^{RL}}_{3x}(t,X)\big]\,,
\end{equation}
\[B^{^{RL}}_{3x}(t,X)=2\,A^{^{RL}}_{1x}(t,X)\,, \ \
C^{^{RL}}_{3x}(t,X)=D^{^{RL}}_{1x}(t,X)\,, \ \
D^{^{RL}}_{3x}(t,X)=C^{^{RL}}_{1x}(t,X)\,.\]

In Fig.~5 we demonstrate the {\it right-left} effects caused by the P-even term (Spqk) in the matrix element squared.

\begin{figure}

\includegraphics[width=0.23\textwidth]{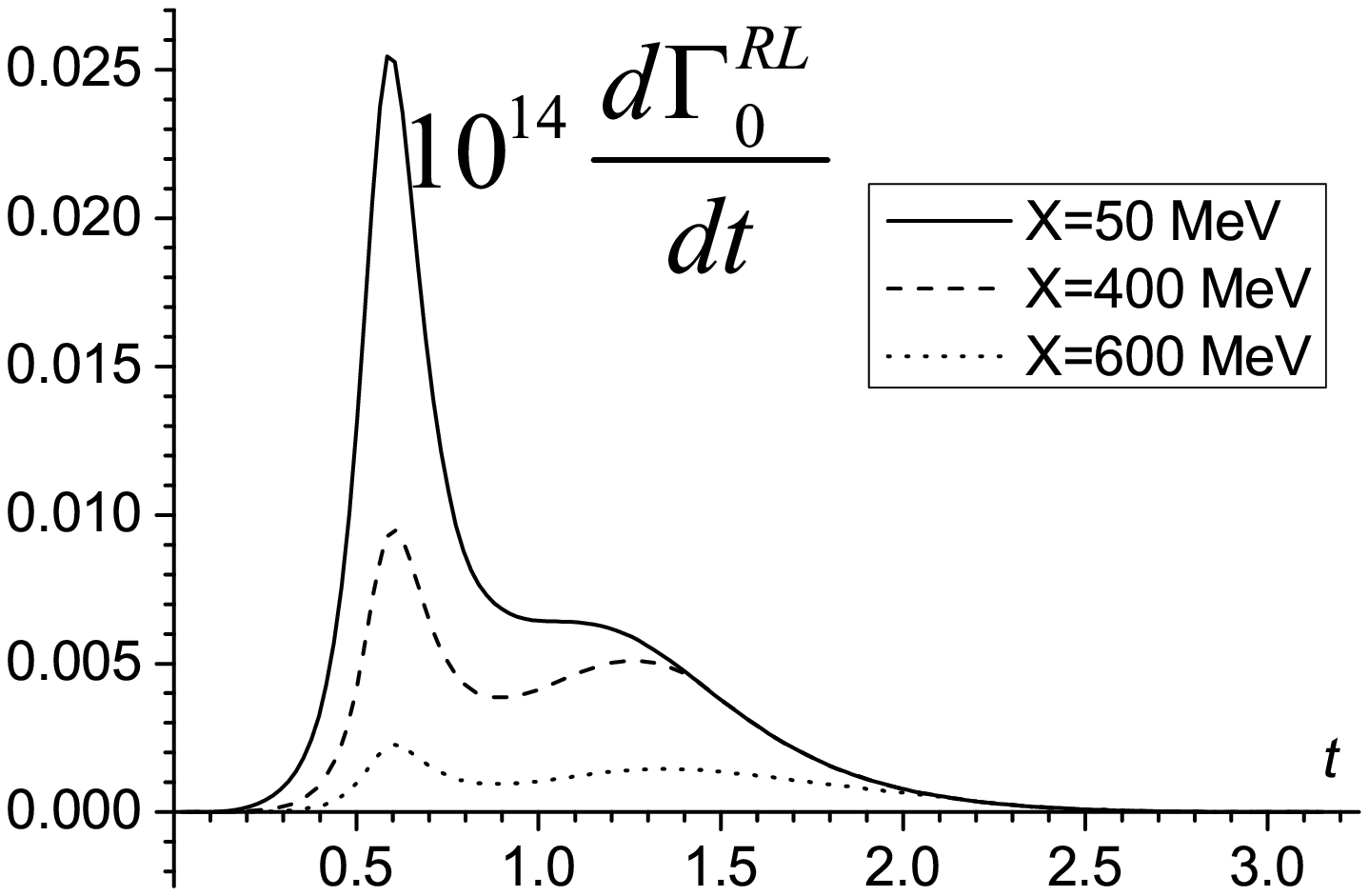}
\includegraphics[width=0.23\textwidth]{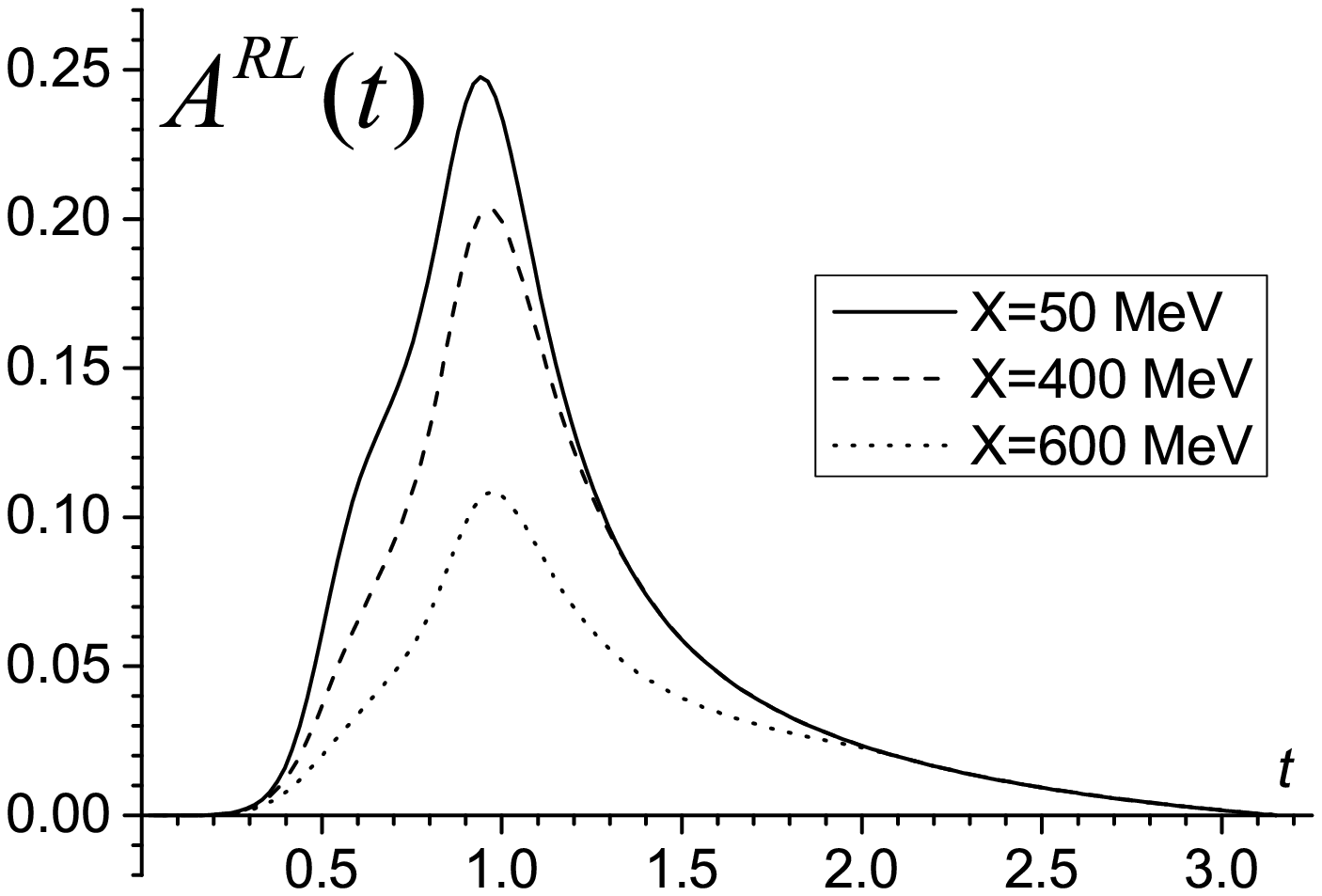}
\includegraphics[width=0.23\textwidth]{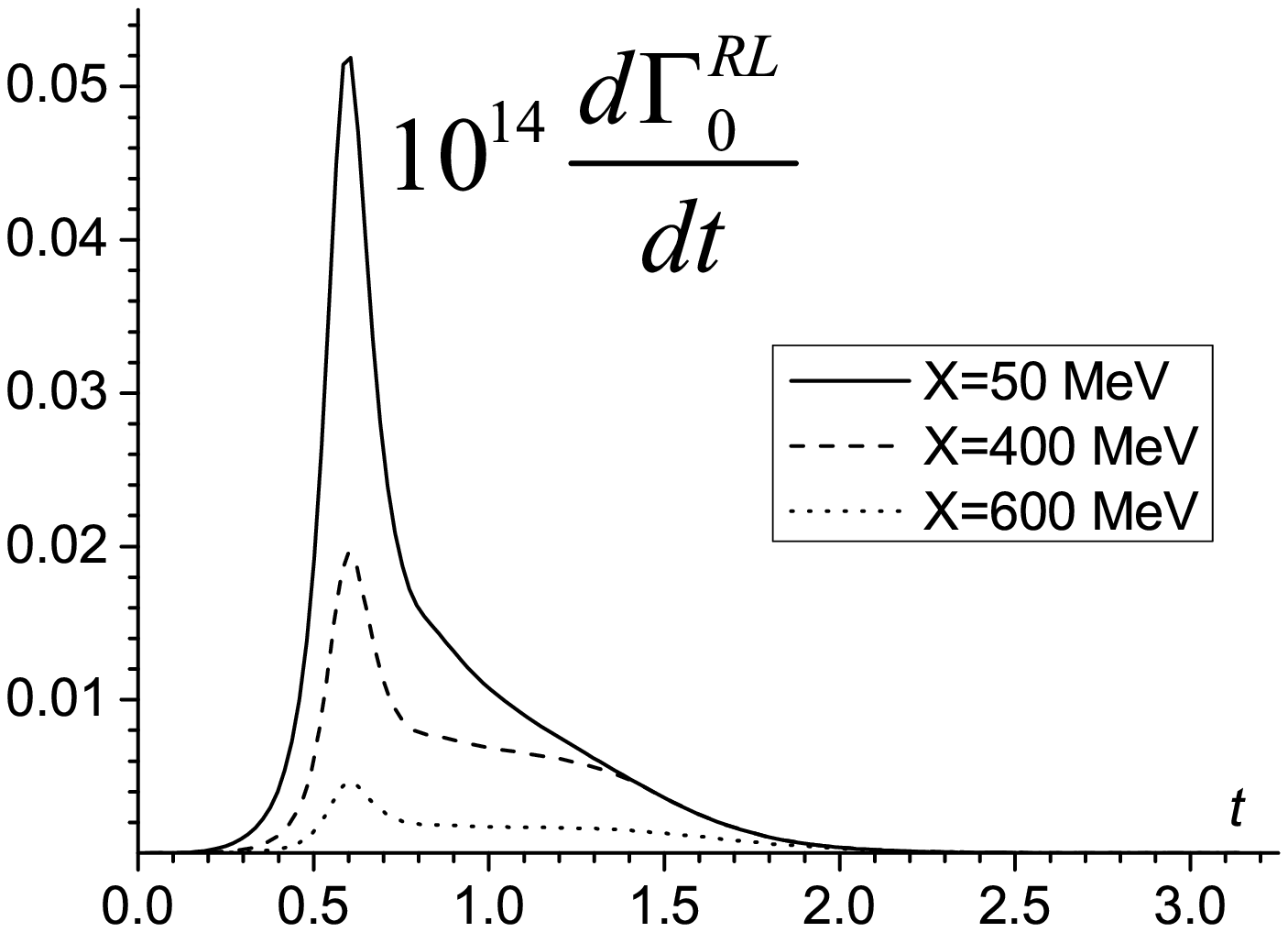}
\includegraphics[width=0.23\textwidth]{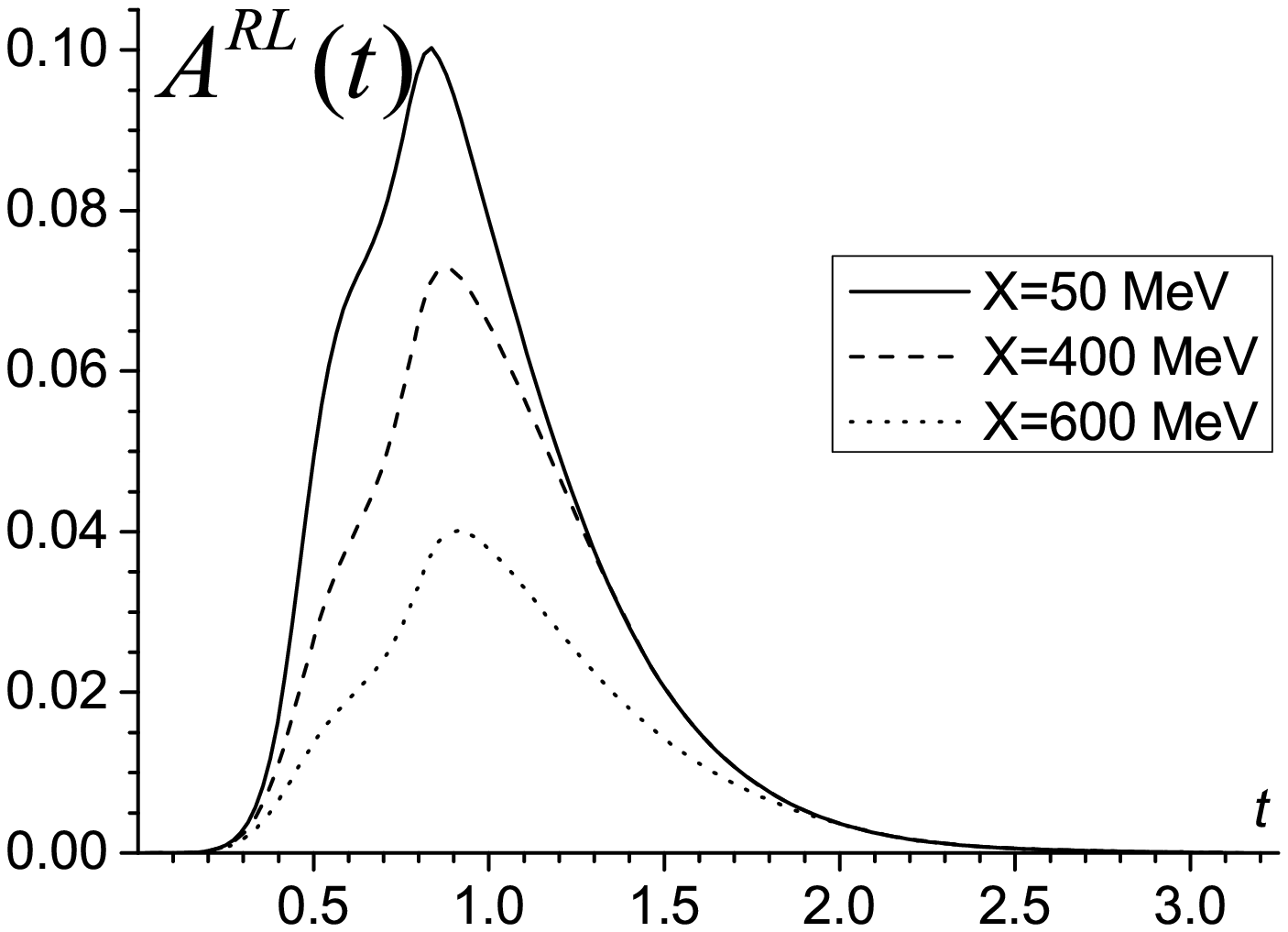}
\vspace{0.4cm}

\includegraphics[width=0.30\textwidth]{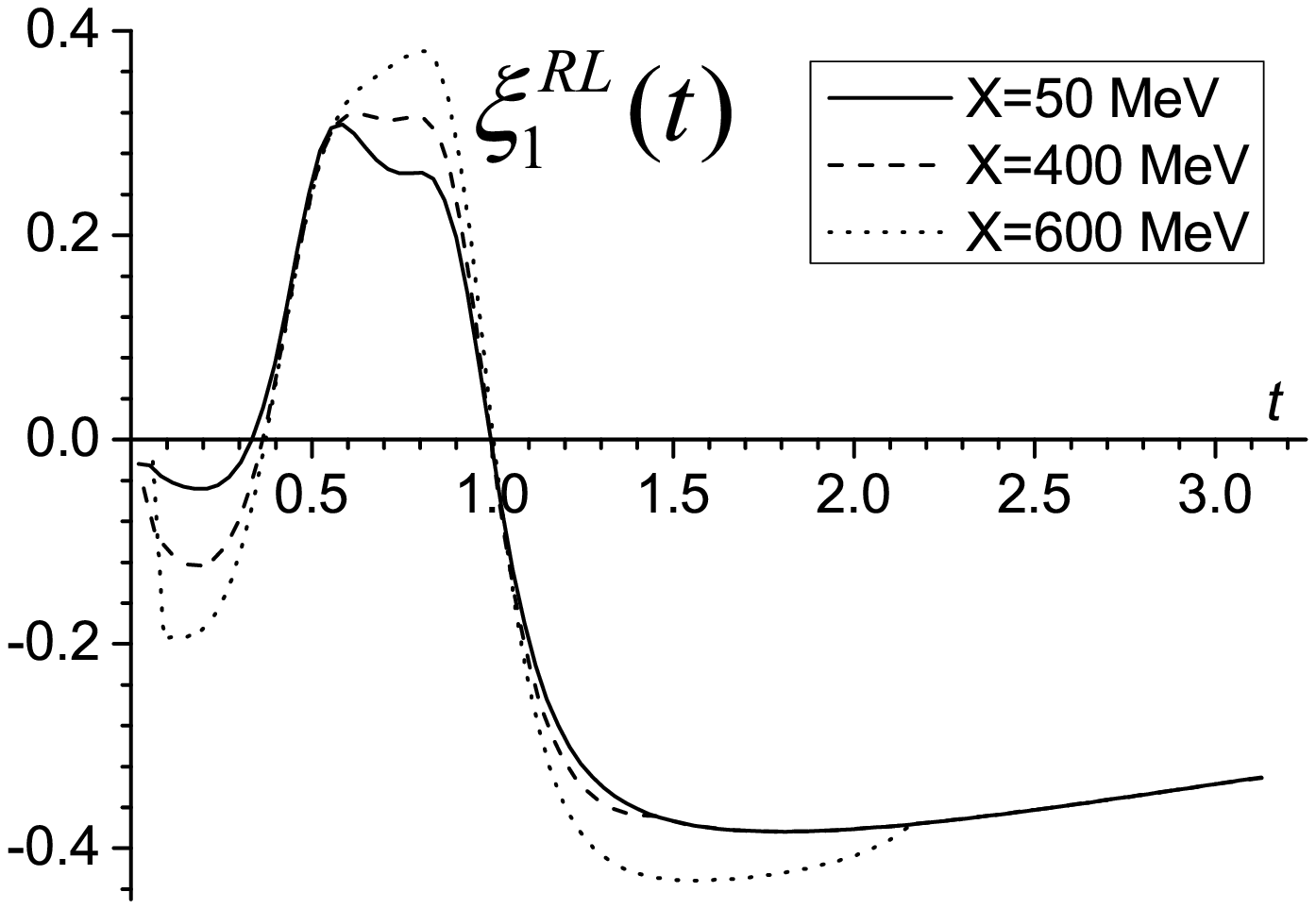}
\includegraphics[width=0.30\textwidth]{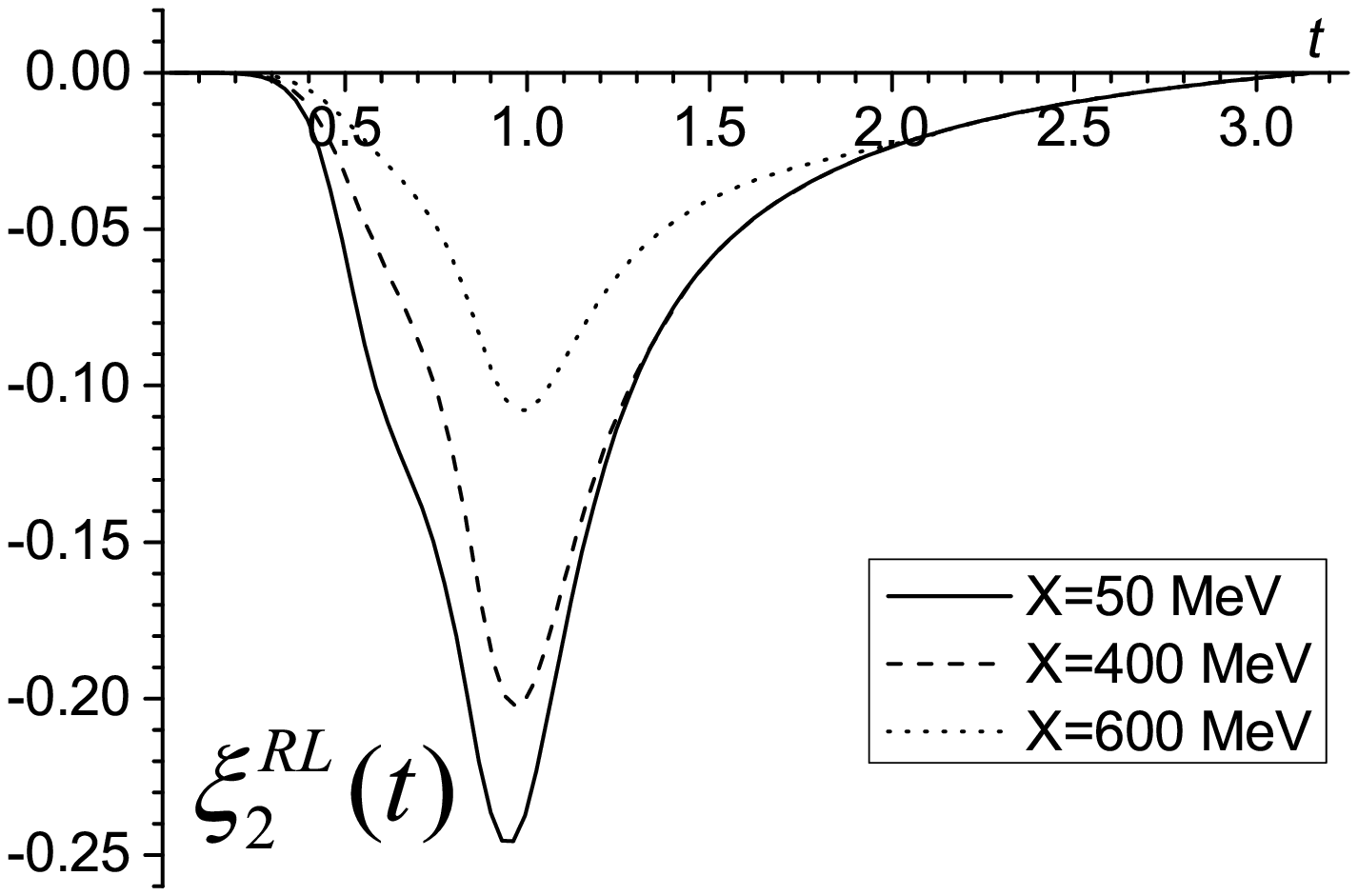}
\includegraphics[width=0.30\textwidth]{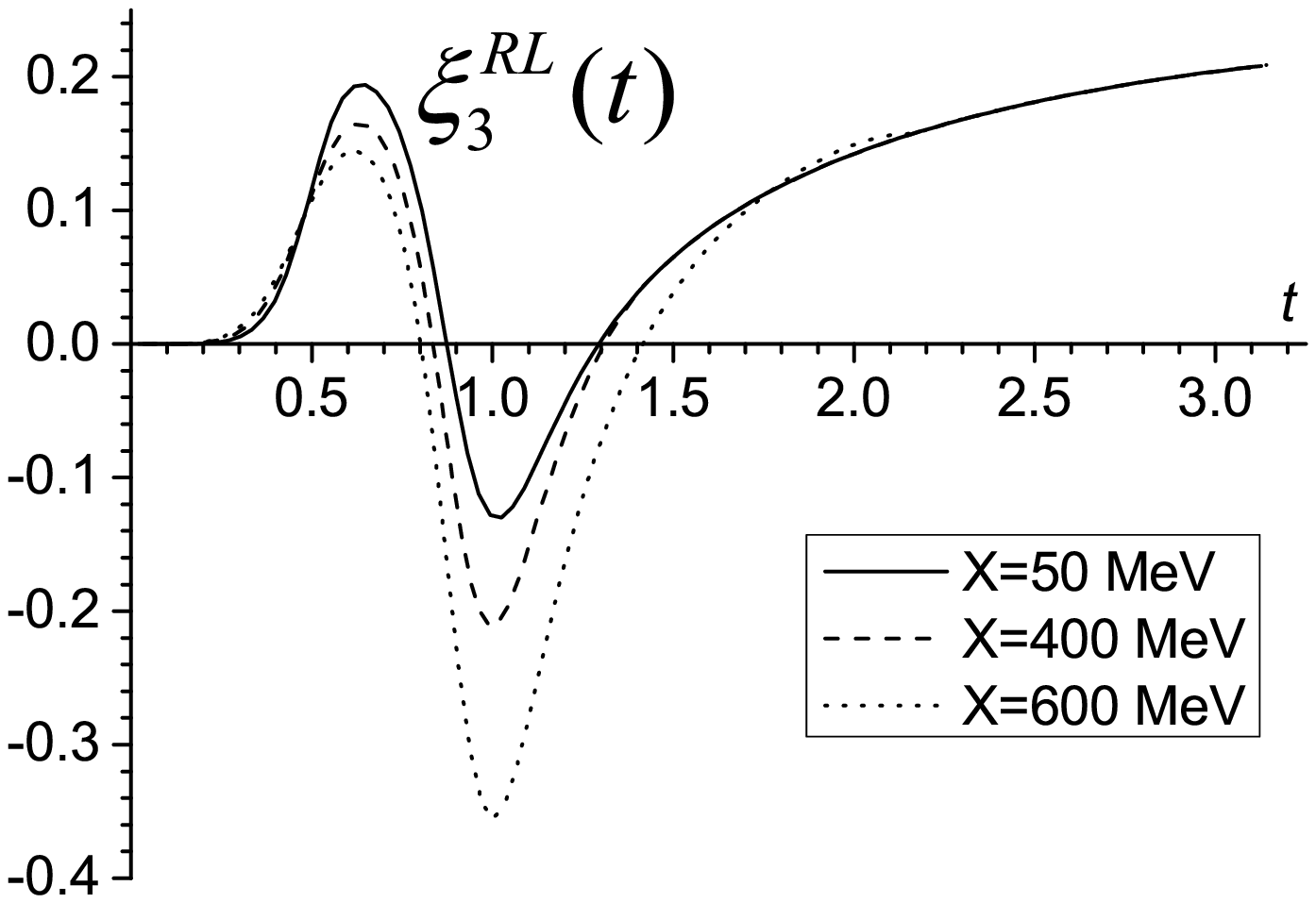}
\vspace{0.4cm}

\includegraphics[width=0.30\textwidth]{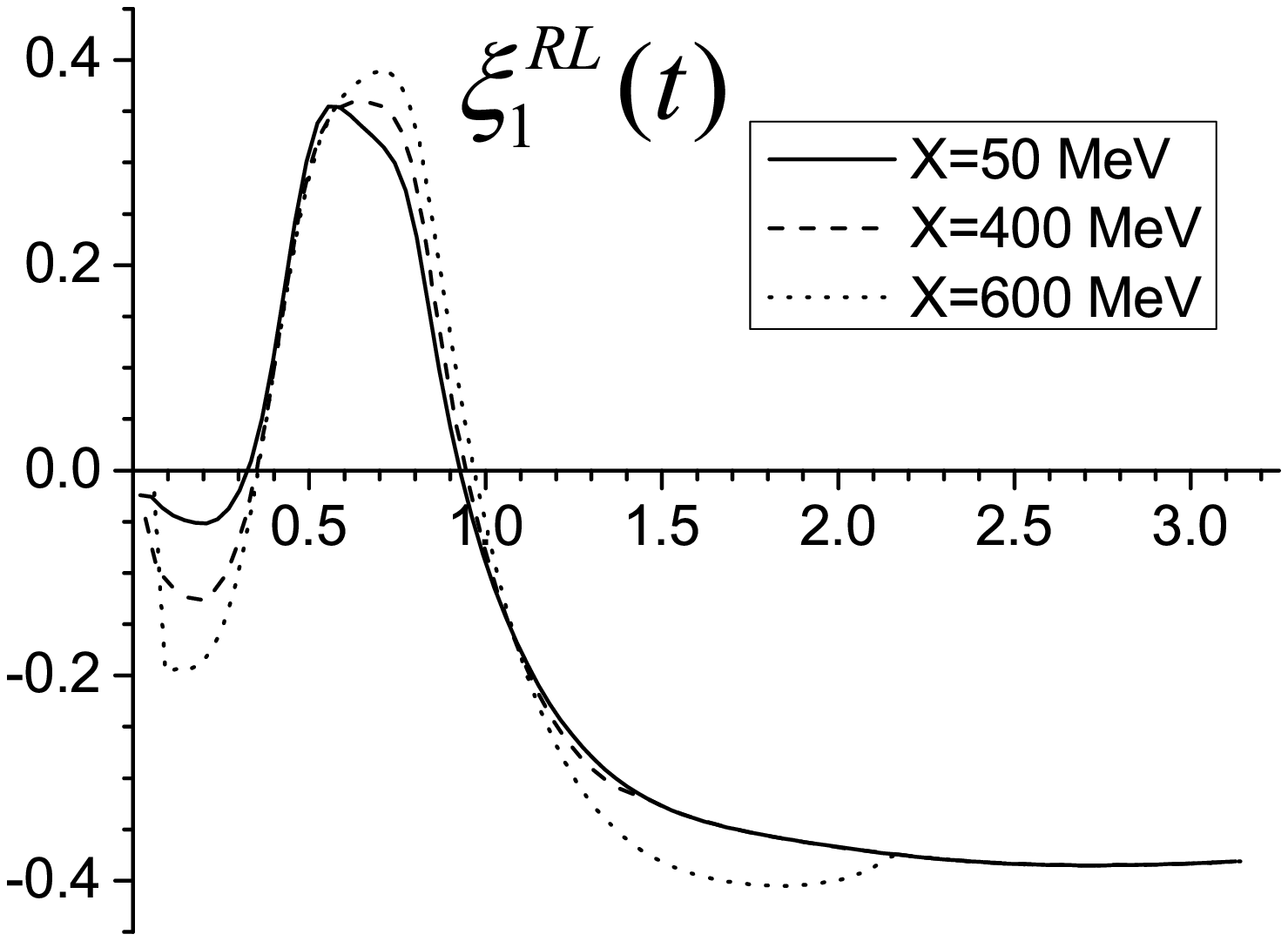}
\includegraphics[width=0.30\textwidth]{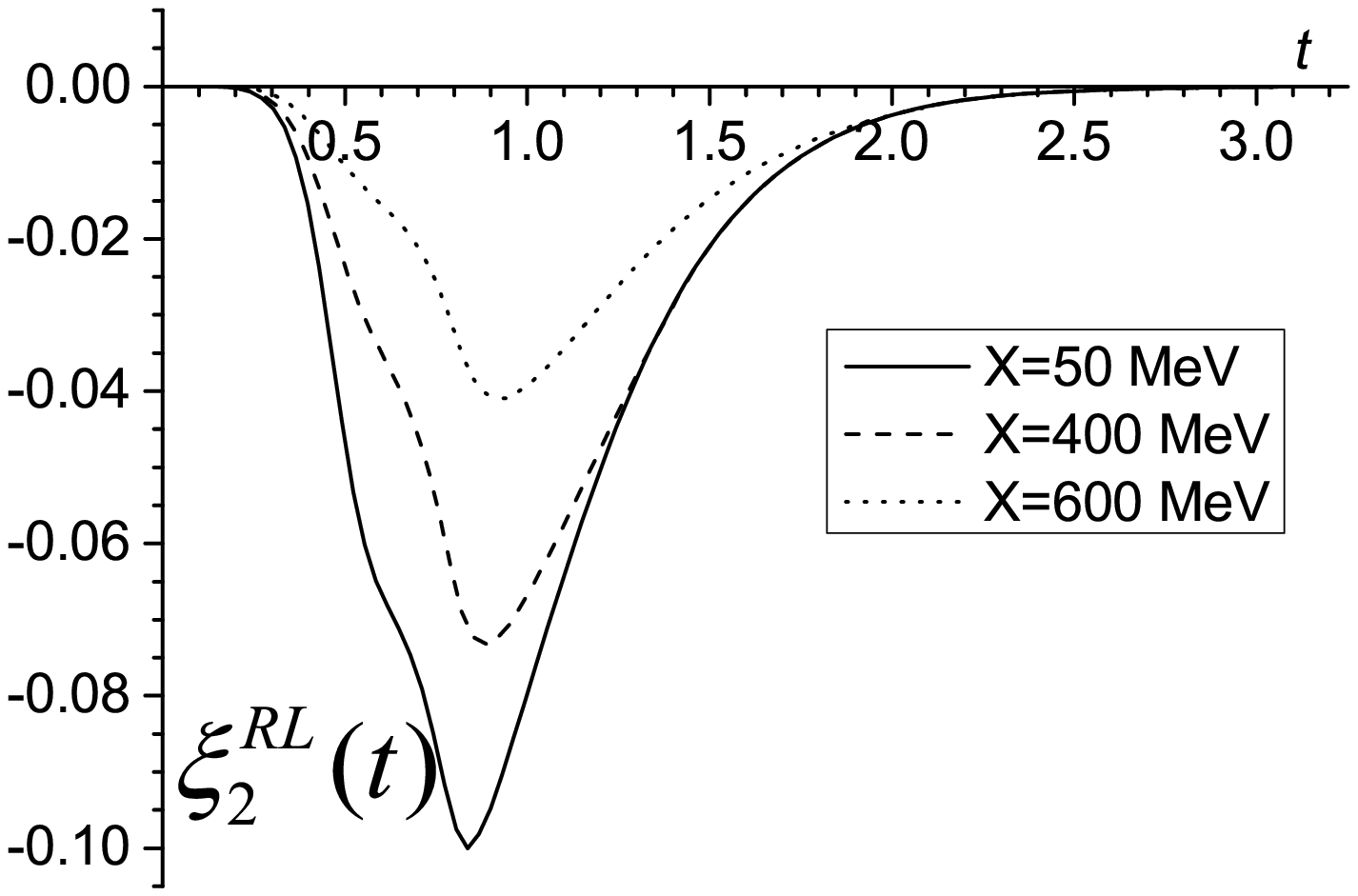}
\includegraphics[width=0.30\textwidth]{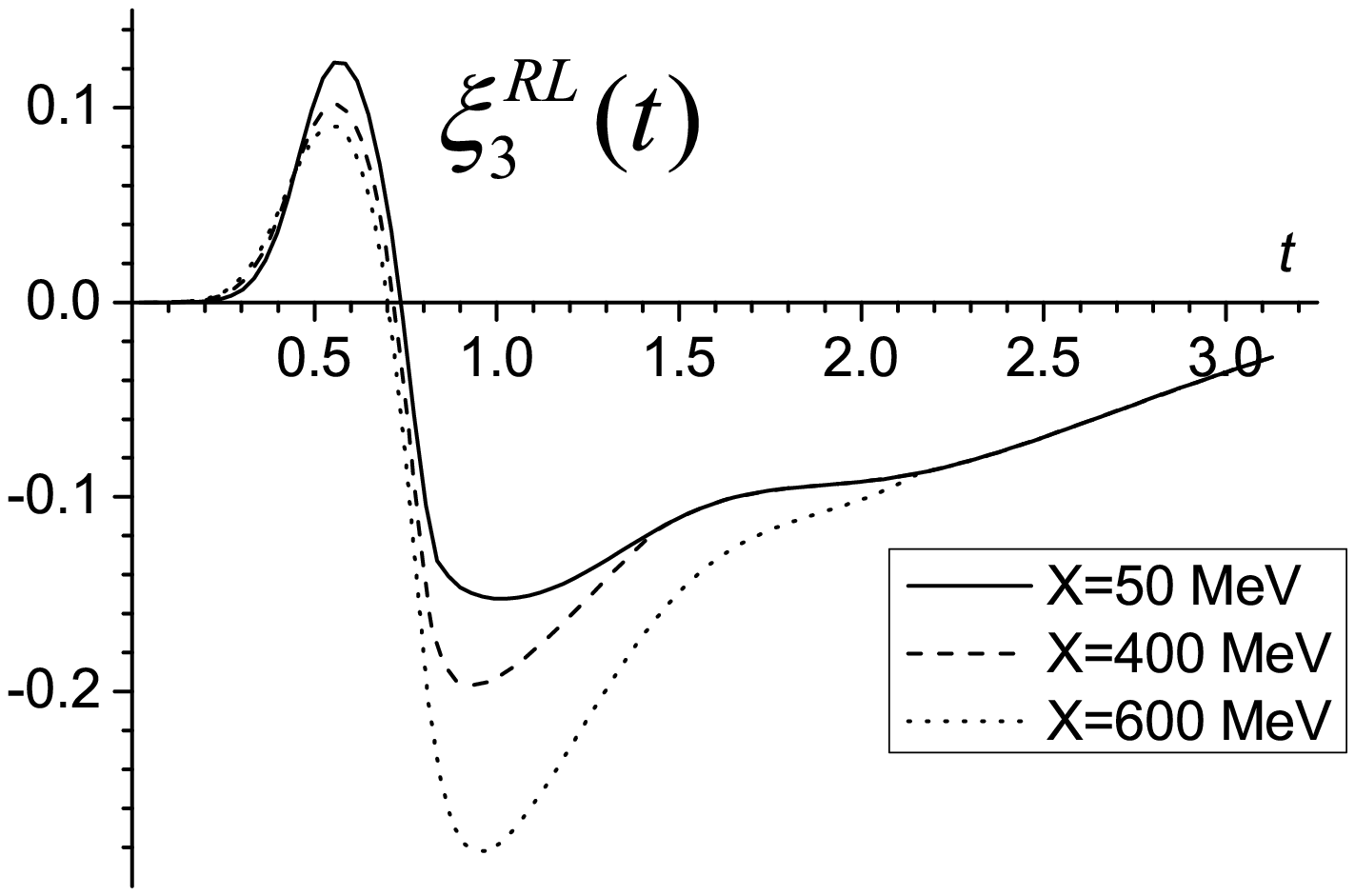}

\caption{The $t-$distribution of the {\it right-left} decay width, the polarization  asymmetry and the correlation parameters defined by Eq.~(\ref{eq:23}). The panels are the same as for the {\it up-down} case . All these quantities arise due to the P-even term (Spqk) in the matrix element squared.}
\end{figure}

\section{Conclusion}

\hspace{0.7cm}

The analytical expressions for the partial differential width and some polarization observables (the right-left and up-down asymmetries) in the process (\ref{eq:1}), as a function of the pion-photon invariant mass squared $t$ and the value of the photon-energy cut X, have been obtained in terms of the $t-$dependent weak vector and axial-vector form factors. All our formulae, with the obvious changes of the masses $m\to m_K$ and values entering in the constant $Z: \, F_\pi\to F_K, \, V_{ud}\to V_{us},$ are applicable for the analysis of the decay $\tau^{\pm}\to K^{\pm}\,\nu_\tau\,\gamma.$ In the last case, of course, one has to use another expressions for the $t-$ dependent form factors (see, for example, \cite{D93, G10}). Our formulae are universal and can be suitable to test different theoretical approaches and models used for the description of
 the $t-$dependence of the corresponding form factors. Among them are the modifications of the vector dominance model \cite{D93, R95}, the light-front quark model \cite{G03}, different parameterizations of the chiral effective theory with resonances \cite{G10, GKKM15, EGPR89}. The interesting one is the Nambu-Jona- Lasinio model and its extended version, which have been widely applied to different semileptonic $\tau$ decays (see \cite{VA17, VNP19, VP18, VPO18} and references therein).

Our calculations indicate large sensitivity to the  parametrization of the form factor such quantities as the differential decay widths $d\,\Gamma_0(t), \ d\,\Gamma_0^S(t), $ and $d\,\Gamma_0^{RL}(t)$, the Stokes parameter $\xi_1(t),$ the {\it right-left} asymmetry $A^{RL}(t)$ as well the correlation parameters
 $\xi_1^{ud}(t), \ \xi_2^{RL}(t)$ and $\xi_3^{RL}(t).$

It was shown the strong dependence of the separate contributions (IB, Resonance and Interference) on the X value. In the region $t\geq 0.5$ GeV$^2$, the resonance contribution dominates and the relative contribution as compared with IB one increases essentially when photon-energy cut becomes greater. Meanwhile, the total contribution, i. e., the t-distribution, decreases more slowly.

The Stokes parameters of the emitted photon, as a function of the variable t, show maximum (at $t\sim 0.5$ GeV$^2$) and minimum (at $t\sim 1$ GeV$^2$) for the $\xi_1$ and $\xi_2$. The increasing of the photon-energy cut X leads to the decreasing of the maximum and increasing of the minimum for the $\xi_1$ parameter. The dependence of the $\xi_2$ parameter on the variable X is opposite to the $\xi_1$ behaviour. The sensitivity of the Stokes parameters to the value X is essentially weak beyond the maximum and minimum. The dependence of the $\xi_3$ parameter on the variable X is similar to the $\xi_1$ and $\xi_2$ ones but the maximum and minimum take place at smaller t values.

 The dependence of the up-down decay width and the polarization asymmetry $A^{ud}$, as a function of the variable t, on the photon-energy cut X is sizeable in the region of the maximum. The correlation parameters $\xi_2^{ud}$ and $\xi_3^{ud}$ are weakly dependent on the variable X. The parameter $\xi_1^{ud}$ is essentially dependent on the variable X in the region of the minimum.

The right-left decay width $d\Gamma_0^{RL}/dt$ and the polarization asymmetry $A^{RL}$ have strong dependence on the variable X in the region of the maximum. The correlation parameter $\xi_1^{RL}$ is weakly dependent on the variable X, but the correlation parameters $\xi_2^{RL}$ and $\xi_3^{RL}$ are strongly dependent on the variable X in the region of the minimum ($t\sim 1$ GeV$^2$).

\section{Acknowledgments}
This work was partially supported by the Ministry of Education and Science of Ukraine
(projects no. 0117U0004866 and no. 0118U002031).
%The research is carried on in the frame of the France-Ukraine IDEATE International Associated Laboratory (LIA).

\end{document}